\immediate\write18{makeindex -s nomencl.ist -o "\jobname.nls" "\jobname.nlo"}
\documentclass[a4paper]{article}
\pdfoutput=1
\usepackage[utf8]{inputenc}
\usepackage[round]{natbib}
\usepackage{amsmath,amsthm,amssymb}
\usepackage{graphicx}
\usepackage{rotating}
\usepackage{setspace}
\usepackage{subcaption}
\usepackage{fullpage}
\usepackage{authblk}
\usepackage{lineno}
\usepackage{hyperref}
\usepackage{xcolor}
\usepackage{nomencl}
\hypersetup{
    colorlinks=true,
    citecolor=blue,
}
\makenomenclature

\begin{document}
\title{Physical observations of the transient evolution of the porosity distribution during internal erosion using spatial time domain reflectometry}
\author[1]{Adnan Sufian\footnote{Corresponding Author, email: \href{a.sufian@uq.edu.au}{a.sufian@uq.edu.au}, ORCiD: 0000-0003-2816-3250}}
\author[1,2]{Tilman Bittner}
\author[1]{Thierry Bore\footnote{ORCiD: 0000-0001-6084-094X}}
\author[1,3]{Mathieu Bajodek\footnote{ORCiD: 0000-0001-8562-5031}}
\author[1]{Alexander Scheuermann} 
\affil[1]{School of Civil Engineering, The University of Queensland, Brisbane, Australia}
\affil[2]{Federal Railway Authority, Germany}
\affil[3]{Laboratoire d'analyse et d'architecture des systèmes (LAAS-CNRS), Université de Toulouse, Toulouse, France}
\date{}
\setcounter{Maxaffil}{0}
\renewcommand\Affilfont{\itshape\small}

\maketitle
\renewcommand\nomname{List of Notations}
\setlength{\nomlabelwidth}{1.5cm}

\begin{abstract}
A purpose-built permeameter was used to explore the transient evolution of porosity during the mixing process in filtration experiments.
The experiments considered upward seepage flow and explored the influence of base and filter particle sizes, along with different hydraulic conditions. 
The permeameter acted as a coaxial transmission line enabling electromagnetic measurements based on spatial time domain reflectometry, from which the porosity profile was obtained using an inversion technique.
Quantitative characteristics of the onset and progression of the mixing process were extracted from a porosity field map.
The limiting onset condition was influenced by geometric and hydraulic factors, with the critical flow rate exhibiting a strong dependence on the base particle size, while the critical hydraulic gradient exhibited a stronger dependence on filter particle size.
The progression of the mixing process was characterised by both the transport of base particles into the filter layer, as well as the settlement of the filter particles into the base layer due to the reduction of the effective stress at the base-filter interface leading to partial bearing failure.
The rate of development of the mixture zone was strongly dependent on the hydraulic loading condition and the base particle size, but the final height of the sample after complete mixing was independent of the hydraulic loading path.
\end{abstract}

\setlength{\parindent}{0em}
\setlength{\parskip}{1em}

\section{\label{sec:introduction}Introduction}

Internal erosion occurs when soil particles within the body of a dam or its foundations are detached and transported under seepage flows.
Internal erosion is attributed to approximately half of dam failures globally \citep{foster2000}.
A key design element to mitigate internal erosion are filters, which comprise coarser granular materials placed adjacent to finer soil, termed the base material.
Modern filter design aims to specify the range of particle size distributions of the filter material so as to prevent the erosion of the finer base material (i.e. the retention function), whilst maintaining sufficient permeability (i.e. the drainage function) \citep{icold2017}.
However, existing dams may locally have filters that do not conform with current design practices, for which conditions \citet{foster2001} proposed 'no erosion', 'excessive erosion' and 'continuing erosion' boundaries.
For an ineffective filter, such as the case for a filter that results in continuing erosion, the transport of the base particles leads to a transient evolution of the porosity distribution within the filter.
This can lead to cavities within the dam structure and can influence the retention and drainage function of the filter.
Existing approaches to measure the transient evolution of the filtration process are typically based on macroscopic observations and point measurements in laboratory experiments.
This paper demonstrates how physical observations can be obtained using spatial Time Domain Reflectometry (spatial TDR), which can provide information on the transient evolution of the local porosity distribution as base material is transported into the filter resulting in a mixture of both materials.
The onset and progression of internal erosion is yet to be fully understood \citep{bonelli2012} and the application of spatial TDR aims to provide new insights into the internal erosion process, which would otherwise not be possible in other physical experiments when only considering macroscopic observations and point measurements.

In this study, the transport of base material into the filter under seepage flow perpendicular to the base-filter interface will be termed filtration, which will be the focus of this paper. 
The term contact erosion is typically used to describe the internal erosion process when flow is parallel to the base-filter interface \citep{beguin2012}.
The filter effectiveness is assessed using geometric criteria based on characteristic particle sizes of the base and filter materials, including the well-known Terzaghi criterion \citep{terzaghi1948,fannin2008}.
The \citet{foster2001} geometric criteria for 'no erosion', 'excessive erosion' and 'continuing erosion' boundaries are also based on characteristic particle sizes.
While the Terzaghi criterion is limited to uniform soils, \citet{ziems1969} provided an alternate criterion for base and filter materials with a wider coefficient of uniformity.
In addition, hydraulic criteria may also be considered, particularly when the geometric criteria in filter design is not satisfied in existing structures.
For the case of contact erosion, \citet{brauns1985} identified the size ratio of base-filter combinations where the geometric and hydraulic conditions governed the erosion process.
Depending on the direction of the flow relative to the base-filter interface, different values for critical hydraulic gradients have been proposed \citep{zweck1959,ziems1969,wittmann1982}. 

A central element of the mixing process in filtration experiments is the transient evolution of the porosity distribution. 
Changes in the porosity distribution are also indicative of changes in the conductivity of the soil. 
Hence, measuring the change in porosity during filtration experiments provides a quantifiable method to assess the onset and progression of erosion.
Different methods have been considered to determine the porosity distribution in physical laboratory experiments, including observation of changes in layer heights \citep{ke2012} and washed out particles \citep{ke2014,rochim2017}, use of gamma rays \citep{alexis2004,sibille2015}, computed tomography scans \citep{homberg2012,sufian2013,fonseca2014} and electromagnetic methods using spatial TDR \citep{scheuermann2010,scheuermann2012}. 

While all of these methods have their respective limitations and advantages, only spatial TDR is able to determine the transient evolution of the porosity distribution with good accuracy and sufficient temporal and spatial resolution.
Conventional TDR systems simply measure the propagation velocity of an electromagnetic signal travelling through a transmission line embedded in the material to be characterised. 
The velocity of this signal is primarily a function of the electrical permittivity of the material.
TDR techniques have been extensively applied to monitor the water content in soils \citep{santamarina2001}, where the permittivity measurements are related to water content through empirical models \citep{topp1980,dirksen1993} or theoretical mixing equations \citep{sihvola1999,bore2018}.
However, conventional TDR is restricted to point-wise measurements or averaged measurements along the transmission line, thereby requiring a sufficient number of TDR probes to obtain spatial distributions.
\citet{scheuermann2012} demonstrated that the TDR trace, especially the section between the first and second main reflection, contains information enabling the determination of the permittivity profile from the measured TDR signal, either in time domain or frequency domain.
The classical approach to achieve this is based on the calculation of the wave propagation along the transmission line due to an incident voltage step \citep{lundstedt1996,norgren1996,feng1999,heimovaara2004,greco2006,bumberger2018}. 
An alternative approach was proposed by \citet{schlaeger2005} which allows the fast computation of soil moisture profiles along elongated probes from TDR signals, using either one- or two-ended measurements. 
This approach is termed spatial TDR and has been extensively used in combination with flat ribbon cables for large-scale applications including flood levees \citep{scheuermann2009} and coastal sand dunes \citep{fan2015}.
In addition, spatial TDR has been applied to laboratory experiments to investigate pressure profile measurements \citep{scheuermann2008} and moisture content in unsaturated soils \citep{yan2021}.

This paper investigates the application of spatial TDR to determine the onset and progression of the mixing process in filtration experiments based on the measured local porosity distribution using a purpose-built coaxial cell permeameter \citep{bittner2019}.
Different size ratio of base and filter materials are considered, with a focus on base-filter combinations that meet the continuing erosion criteria by \citet{foster2001}. 
In addition, different hydraulic boundary conditions are considered to investigate the influence of the rate of hydraulic loading.
Analysis of the spatial and temporal evolution of the local porosity distribution provided new insights into factors that contribute to the onset and progression of mixing in filtration experiments.

\section{\label{sec:experiments}Experimental Apparatus and Program}

\subsection{\label{sub:coaxial-erosion-cell}Coaxial Erosion Cell}

\begin{figure}[ht]
    \centering
    \includegraphics[page=1, trim=1cm 18cm 5cm 0cm, clip]{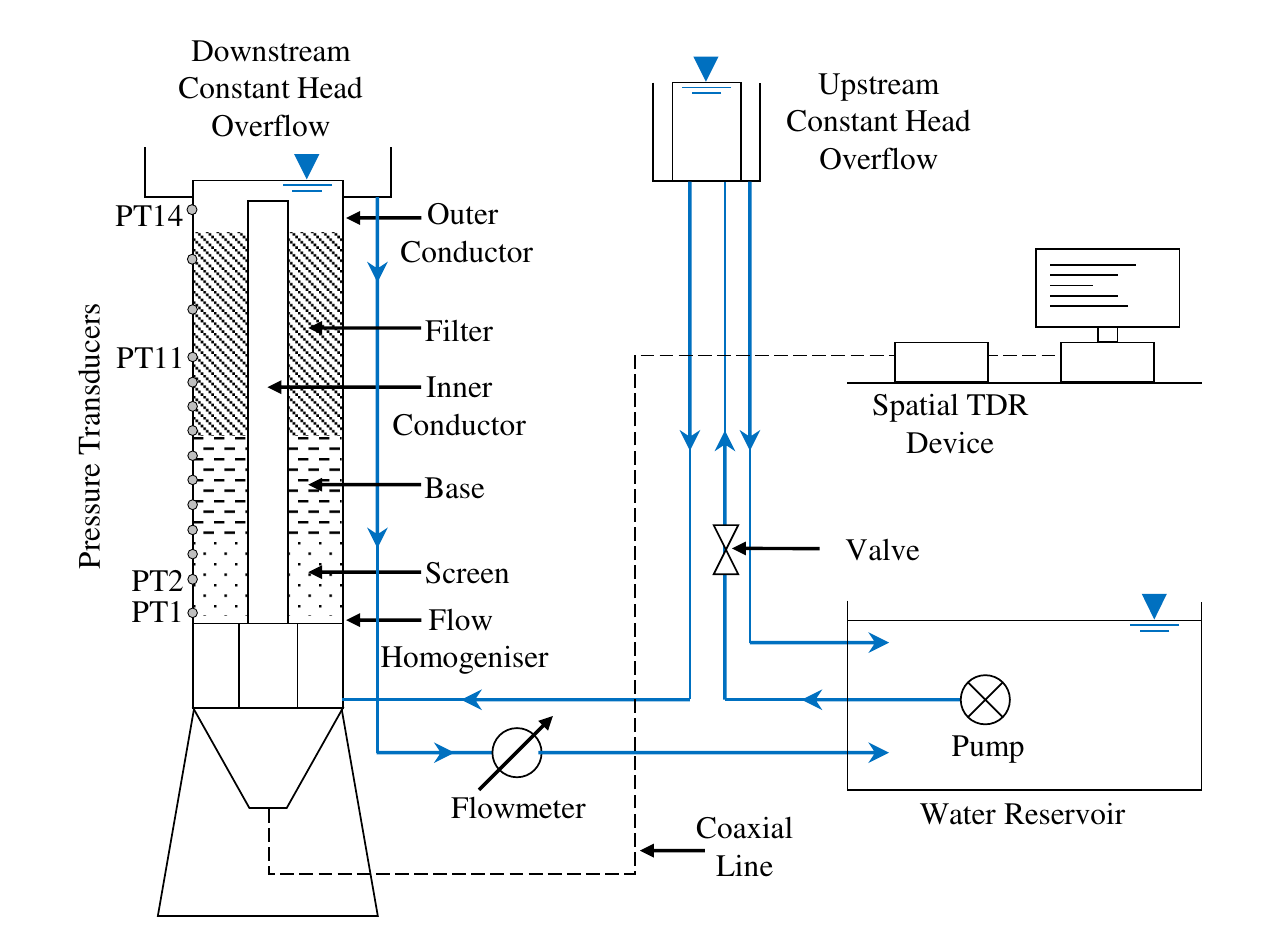}
    \caption{\label{fig:coaxial-cell}Schematic diagram of the experimental apparatus (not to scale).}
\end{figure}

A laboratory permeameter enhanced with capabilities for investigating the spatial and temporal evolution of the porosity distribution using spatial TDR was designed and constructed.  
The experimental apparatus consists of a coaxial erosion cell, a closed water cycle system and various instrumentation.
A schematic of the experimental setup is provided in Figure~\ref{fig:coaxial-cell}. 
A detailed description of the experimental apparatus (including calibration and validation) has been presented in \citet{bittner2019}. 
A brief summary of the apparatus is presented below, as this paper focusses on the application of the coaxial erosion cell to investigate the mixing process in filtration experiments.

The coaxial erosion cell is a copper-built permeameter constructed as a coaxial transmission line in order to enable electromagnetic measurements from which the longitudinal porosity distribution along the sample could be obtained.
A sample of up to 45 cm in length can be setup within the annulus formed by the inner and outer conductors of the coaxial erosion cell.
The inner conductor has an outer diameter of 41.3 mm and the outer conductor has an inner diameter of 151.9 mm.
An observation window 4 cm wide and 42 cm in height enables visual monitoring of the erosion process on the outer surface of the cell.
The base of the observation window is the reference datum from which measurements of heights are provided.
The geometry of the coaxial erosion cell was chosen as it provided an optimised design where the electric field was perfectly bounded between the inner and outer conductor.

Electromagnetic measurements along the coaxial erosion cell were obtained using a time domain reflectometer capable of generating a voltage step and recording the reflected response.
The porosity distribution along the cell could be determined from the TDR trace using an inversion technique.
A summary of the inversion technique is presented below, which highlights the three key steps. 
The first step consists in computing the spatially distributed transmission line parameters. 
Based on the assumption of a pure Transverse Electromagnetic Mode propagation along the transmission line, the propagation of an electromagnetic signal can be described by the telegrapher's equation. 
In this framework, relevant properties can be described by lump element circuits \citep{pozar2011}, where the inductance and resistance depends on the transmission line and are assumed to be constant, whereas the capacitance and conductivity depends on the material and are assumed to be spatially dependant. 
It was assumed that that the resistance is zero and that the conductivity is constant, and hence, the objective is to determine the capacitance profile.
Using the algorithm developed by \citet{schlaeger2005}, the telegrapher's equation was numerically solved with appropriate initial conditions to obtain a simulated TDR trace, $V_o^s(t)$.
The simulated TDR trace is compared with the measured TDR trace, $V_o^m(t)$, using the following objective function:
\begin{equation}
    J\left(C\right) = \sum \int_0^T V_o^s(t,x,C) - V_o^m(t,x) \: \textrm{d}t
\end{equation}
The conjugate gradient method is used to minimise the objective function, by adjusting the capacitance profile, $C(x)$, until both simulated and measured TDR traces sufficiently match. 
The second step comprises computing the dielectric permittivity profile, $\varepsilon(x)$, from the capacitance profile, which is based purely on the geometry of the transmission line (i.e. the coaxial erosion cell) and is given by:
\begin{equation}
    \varepsilon (x) = \frac{C(x) \ln(b/a)}{2\pi}
\end{equation} 
where $a$ and $b$ are the diameter of the inner and outer conductors, respectively. 
The third step is the conversion of the dielectric permittivity profile to a porosity profile, $n(x)$, using theoretical mixing equations.
Prior studies indicated the applicability of the Lichtenecker-Rother model \citep{bittner2017,bittner2019} for saturated glass beads, where the dielectric permittivity of the mixture is computed as the weighted sum of the dielectric permittivity of the water and glass beads by their respective volume fraction:
\begin{equation}
    \varepsilon^a = n\varepsilon_w^a + (1-n)\varepsilon_s^a
\end{equation}
where $\varepsilon_w$ is the permittivity of water, $\varepsilon_s$ is the permittivity of the solid phase, and $a$ is a shape factor.
Experimental investigations suggested that the shape factor is dependent on the soil structure \citep{brovelli2008} and that it should remain as a fitting parameter.
\citet{bittner2019} compared various mixing models and shape factors, and results indicated that a shape factor of  $a = 2/3$ is appropriate.
Further details on the inversion technique is provided in \citet{bittner2019}. 

The closed water cycle system enabled upwards flow in the coaxial erosion cell.
Water entered the base of the cell from an upstream constant head overflow tank. 
The height of the upstream tank is indicative of the applied hydraulic head and could be adjusted to consider various hydraulic boundary conditions, as discussed in Section~\ref{sub:experimental-program}.
At the top of the cell, a constant overflow into a downstream reservoir ensured a constant total head.
All connecting pipes had an inner diameter of 32 mm to minimise the head loss through the pipes.
A 10 mm thick PMMA (polymethyl methacrylate) perforated plate was placed towards the base of the cell, which acted as a flow homogeniser to ensure that a relatively uniform hydraulic boundary condition was imposed upstream of the sample.
The flow homogeniser extended to a height of 2 mm above the base of the observation window, and hence, the sample effectively started from the reference datum set at the base of the observation window.
The flow homogeniser also served an important role for the analysis of the TDR trace. 
The low dielectric permittivity of PMMA resulted in a stiff peak in the TDR trace enabling the start of the sample to be readily determined.

The flow rate was measured using a flowmeter, which was placed downstream of the cell to ensure that its smaller inner diameter did not influence the head loss through the pipes.
The pore water pressure along the cell was measured using 14 pressure transducers, which are marked as PT1 to PT14 in Figure~\ref{fig:coaxial-cell}.
PT1 and PT2 are located 1 cm and 4.5 cm above the base of the observation window.
The pressure transducers from PT2 to PT11 are spaced at 2.5 cm intervals, while they are spaced at 5 cm intervals from PT11 to PT14. 
The hydraulic gradient within the sample was obtained from these point measurements of the pore water pressure.  

\subsection{\label{sub:materials}Materials and Sample Configuration}

Test samples were prepared with soda-lime glass beads as an idealised granular soil.
The suitability and advantages of using glass beads as a replacement material for soil has been highlighted in several prior studies \citep{tomlinson2000,scheuermann2012,zakaraya2015,bittner2017}.
From the perspective of spatial TDR, an additional benefit of considering glass beads is the constant dielectric permittivity of the glass beads, as well as the fact that mineralogy of real soil particles do not have to be taken into account.
Therefore, despite the high sphericity of glass beads and their relatively narrow size distributions, glass beads provided a fundamental basis to explore the mixing process in filtration experiments in a manner not possible in other experimental approaches. 

The sample comprised three zones: a screening layer, the base material and the filter material, as shown in Figure~\ref{fig:coaxial-cell}.
The screening layer consists of an approximately 5 cm layer of 6 mm diameter glass beads, underlying an approximately 4 cm layer of 2 mm diameter glass beads.
The screening layer ensured that base particles did not experience significant downwards penetration into the screening layer, which is further verified by the subsequent analysis of the spatial TDR data.
For the base and filter materials, two different diameter glass beads were considered for each zone.
The base layer was approximately 10 cm in height and consisted of either: (i) Base I, with particle sizes ranging from 0.425 - 0.600 mm; or (ii) Base II, with particle sizes ranging from 0.300 - 0.425 mm. 
The filter layer was approximately 20 cm in height and consisted of either: (i) Filter A, with particle sizes of 8 mm; or (ii) Filter B, with particles sizes of 6 mm.
All glass beads are approximately monodisperse with the exception of those used for the base layer, which exhibited a narrow size range due to the manufacturing process. 

\subsection{\label{sub:experimental-program}Experimental Program}

The four different base-filter combinations formed with the two base and filter materials outlined in Section~\ref{sub:materials} were tested under two hydraulic boundary conditions:
\begin{enumerate} 
\item H1: hydraulic head was increased 1 cm every 10 minutes; or 
\item H2: hydraulic head was increased 1 cm every 5 minutes until the onset of the mixing process, and then subsequently increased in increments of 1 cm and held constant so long as the mixing process was still progressing.
\end{enumerate} 
For both hydraulic boundary conditions, the end of each test was marked by the breakthrough and fluidisation of the base material through the filter material.
For H2, the onset and progression of the mixing process was visually assessed through the observation window. 
While both the onset and progression can be observed for both hydraulic boundary conditions, the subsequent analysis will demonstrate that H1 provides greater detail on the onset, while H2 provides detailed insights into the progression, particularly with respect to equilibrium states of mixing between the base and filter materials.

The complete list of the experimental program is shown in Table~\ref{tab:exp-program}.
The naming convention for each test is $\textrm{B}x\_\textrm{F}y\_\textrm{H}z$, where \emph{x} can take on a value of I or II to signify which base material is considered, \emph{y} can take on a value of A or B to signify which filter material is considered and \emph{z} can take a value of 1 or 2 to signify which hydraulic boundary condition (from the above list) is considered. 
In Table~\ref{tab:exp-program}, the size ratio is defined by $d_{15F}/d_{95B}$ following \citet{foster2001}, where $15\%$ (by mass) of the filter particles are smaller than $d_{15F}$ and $95\%$ (by mass) of base particles are smaller than $d_{95B}$.
Given that Filter A and B are monodisperse, $d_{15F}$ is simply the diameter of the filter particles outlined in Section~\ref{sub:materials}.
The base materials (Base I and II) have a narrow distribution, so $d_{95B}$ was determined by assuming a linear particle size distribution (in log-linear space) and interpolating between the diameter ranges stated in Section~\ref{sub:materials}.
The size ratio varied from slightly to significantly above the ratio of $d_{15F}/d_{95B} = 9$ proposed in \citet{foster2001} for the continuing erosion boundary (as well as the limit of $d_{15F}/d_{85B} = 9$ proposed in \citet{sherard1984}).
In this study, it was advantageous to investigate continuing erosion, as the evolution of the porosity distribution is most pronounced for the case of continuing erosion due to the formation of a significant mixture zone between the base and filter particles.
The thickness of the screen ($l_{screen}$), base ($l_{base}$) and filter ($l_{filter}$) layers upon sample preparation is provided in Table~\ref{tab:exp-program}.
Note that the initial thickness of the mixture layer ($l_{mixture}$) is also recorded to demonstrate that the thickness of the mixture layer initially is minimal.
The average porosity of the base ($n_{ave,B}$) and filter ($n_{ave,F}$) are also listed in Table~\ref{tab:exp-program} and described in detail in Section~\ref{sub:porosity}. 

\begin{table}[ht]
    \caption{\label{tab:exp-program}Experimental program using the coaxial erosion cell}
    \centering
    \begin{tabular}{|c|c|c|c|c|c|c|c|c|c|c|}
    	\hline
        Test & Base & Filter & Hydraulic & $l_{screen}$ & $l_{base}$ & $l_{mixture}$ & $l_{filter}$ & $\frac{d_{15F}}{d_{95B}}$ & $n_{ave,B}$ & $n_{ave,F}$ \\
         & & & Boundary & [mm] & [mm] & [mm] & [mm] & & & \\
         & & & Condition & & & & & & &\\
        \hline
        \hline
        BI\_FA\_H1  & I  & A & 1  & 88.0 & 98.5 & 3.5 & 200.0 & 13.6 & 0.38 & 0.39 \\ 
        BI\_FB\_H1  & I  & B & 1  & 87.0 & 99.0 & 1.0 & 196.0 & 10.2 & 0.37 & 0.37 \\ 
        BII\_FA\_H1 & II & A & 1* & 87.0 & 99.0 & 4.0 & 198.0 & 19.2 & 0.36 & 0.38 \\ 
        BII\_FB\_H1 & II & B & 1  & 87.0 & 96.0 & 8.0 & 195.0 & 14.4 & 0.36 & 0.39 \\ 
        \hline
        BI\_FA\_H2  & I  & A & 2  & 88.0 & 99.0 & 3.0 & 195.0 & 13.6 & 0.36 & 0.37 \\ 
        BI\_FB\_H2  & I  & B & 2  & 87.0 & 99.0 & 2.0 & 194.0 & 10.2 & 0.38 & 0.37 \\ 
        BII\_FA\_H2 & II & A & 2  & 86.0 & 97.0 & 3.5 & 201.5 & 19.2 & 0.35 & 0.39 \\ 
        BII\_FB\_H2 & II & B & 2  & 88.0 & 99.0 & 2.0 & 198.5 & 14.4 & 0.36 & 0.38 \\ 
        \hline
    \end{tabular}

    \raggedright
    \footnotesize{* In BII\_FA\_H1, an initial hydraulic head increase of 5 cm was applied, after which the applied head was increased 1 cm every 10 minutes, as per hydraulic boundary condition 1.}
\end{table}

\section{\label{sec:analysis}Analysis of the Transient Evolution of Porosity in Filtration Experiments}

\begin{figure}[ht]
    \centering
    \begin{subfigure}[b]{0.32\textwidth}
        \centering
        \includegraphics[width=5.5cm]{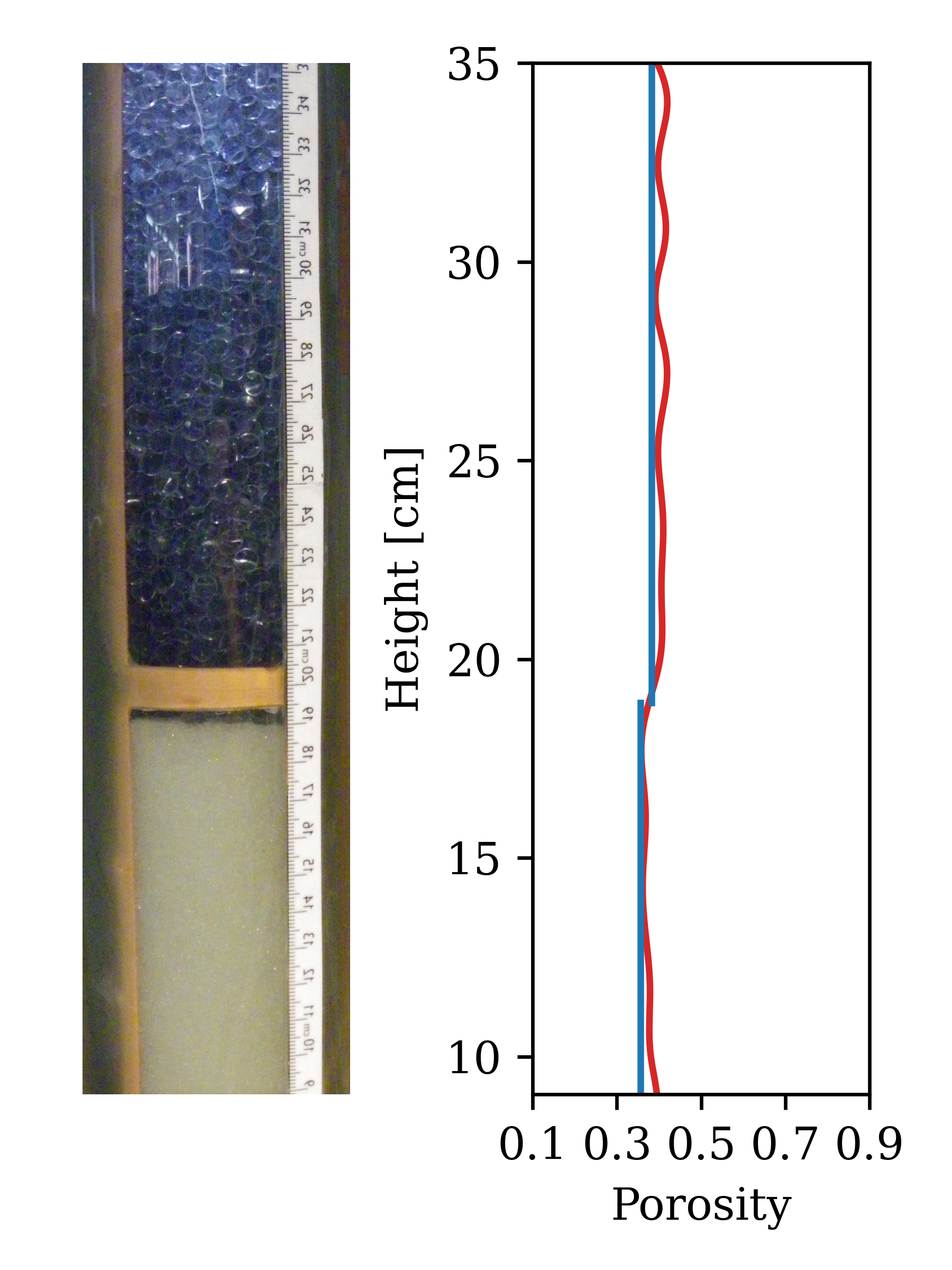}
        \caption{Initial Condition}
    \end{subfigure}
    \hfill
    \begin{subfigure}[b]{0.32\textwidth}
        \centering
        \includegraphics[width=5.5cm]{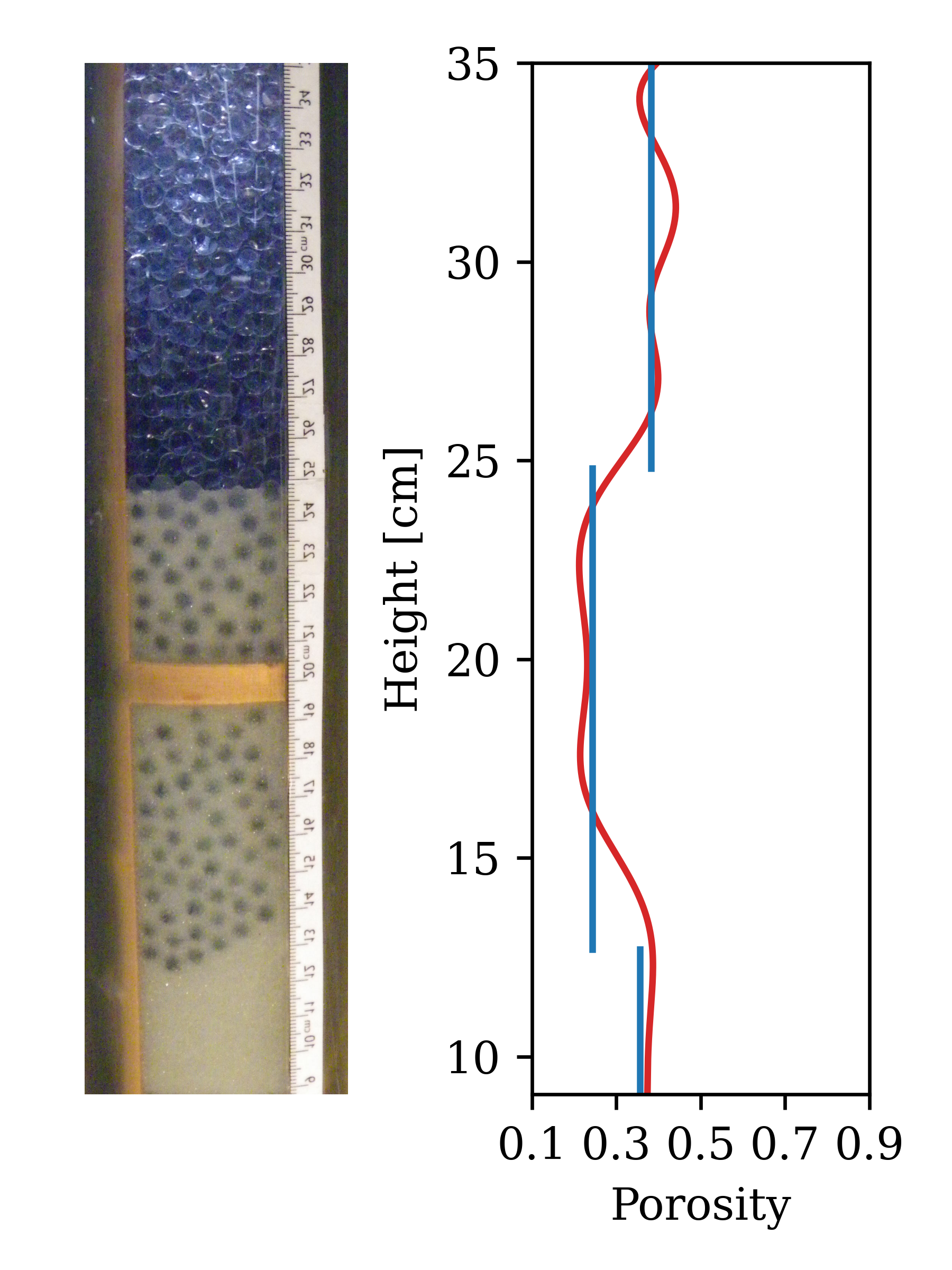}
        \caption{Partial Mixing}
    \end{subfigure}
    \hfill
    \begin{subfigure}[b]{0.32\textwidth}
        \centering
        \includegraphics[width=5.5cm]{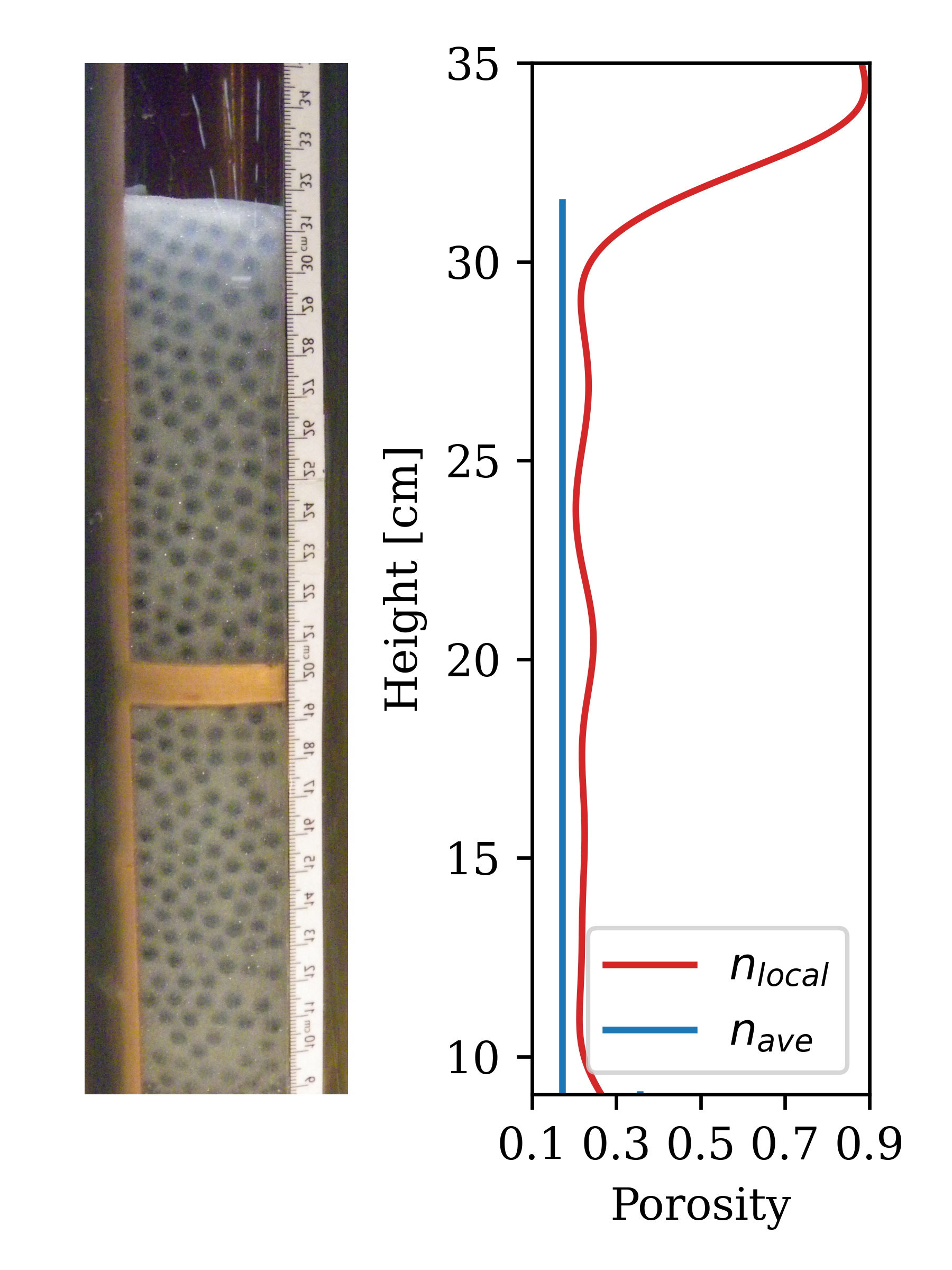}
        \caption{Complete Mixing}
    \end{subfigure}
    \caption{\label{fig:typical-case}Visual representation of the mixing process in filtration experiments is shown for a typical case. The photograph of the coaxial erosion cell at different times illustrates the development of the mixture zone. A comparison between the local porosity profile measured with TDR ($n_{local}$) and average porosity determined by layer heights ($n_{ave}$) is also shown.}
\end{figure}

The key characteristics of the mixing process observed in the filtration experiments are shown in Figure~\ref{fig:typical-case} for a typical case.
Initially, the base and filter materials are distinct layers (Figure~\ref{fig:typical-case}(a)).
Head loss is dominated by the base layer and the hydraulic gradient in the base layer can be calculated from the pressure transducers within the base layer.
As the base and filter layers comprise spherical glass beads with narrow size distributions, minimal differences in the porosity distribution is noted.
A detailed discussion on the average and local porosity distribution is provided below in Section~\ref{sub:porosity}.

\begin{figure}[ht]
    \centering
    \includegraphics[page=1, trim=0cm 0cm 0cm 0cm, clip, scale=0.85]{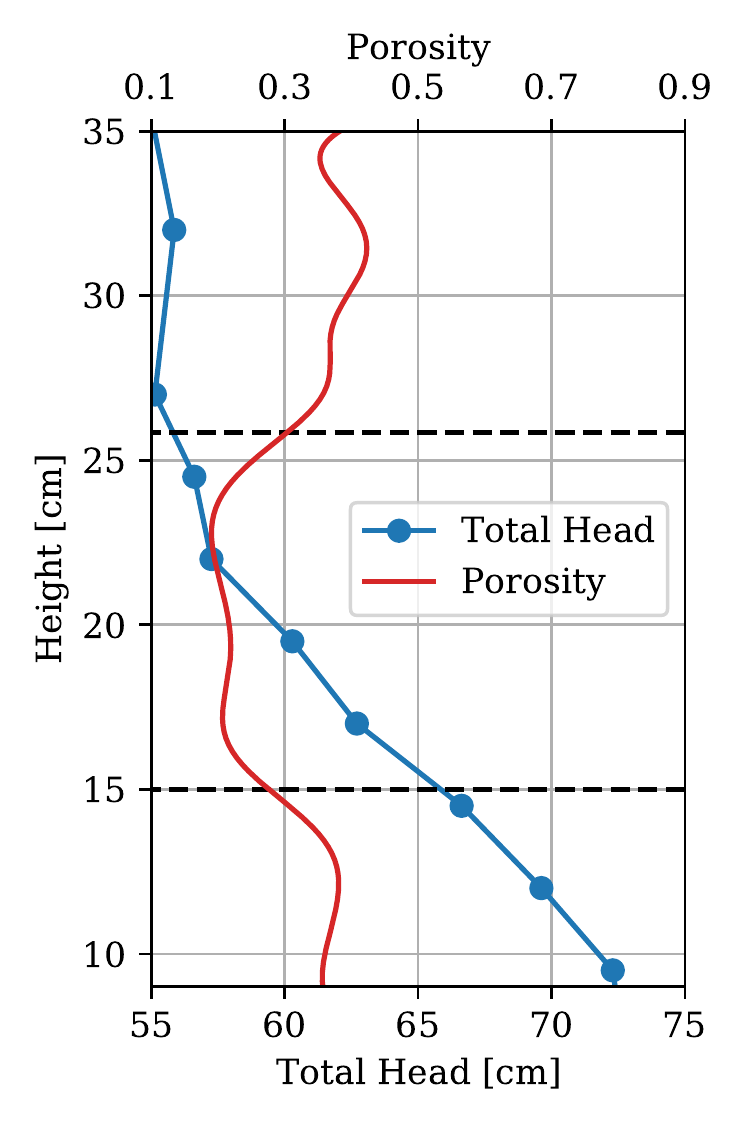}
    \caption{\label{fig:typical-grad}Typical profile of the total head along the length of the sample as measured by pressure transducers. The corresponding local porosity profile measured with spatial TDR is also shown. The mixture zone is indicated by the region bounded by the black dashed lines.}
\end{figure}

Following the onset of erosion, a mixture zone is formed comprising base and filter particles (Figure~\ref{fig:typical-case}(b)).
This is conventionally assumed to be the result of the transport of base particles into the filter under upward seepage flow. 
However, it is also significantly influenced by the settling of the filter particles into the base material. 
Both mechanisms (i.e. transport of base particles and settling of filter particles) were observed in the experiments and are delineated in the analysis detailed in Section~\ref{sub:progression}.
With the formation of a mixture zone, head loss occurs in both the base layer and the mixture zone, as the hydraulic conductivity of the filter material is significantly higher than that of both zones.
A typical profile of the total head when a mixture zone is formed is shown in Figure~\ref{fig:typical-grad}.
As per the initial configuration in Figure~\ref{fig:typical-case}(a), the hydraulic gradient in the base layer can be calculated from the pressure transducers within the base layer, noting that the base layer is contracting as the mixing process progresses. 
The hydraulic gradient in the mixture zone can similarly be determined from the pressure transducers within the mixture zone.
The hydraulic gradient was approximately the same across the thickness of the base layer (as indicated by the linear total head profile), while some spatial variability was noted within the mixture zone, which can be visually observed by the slight curvature of the total head profile shown in Figure~\ref{fig:typical-grad}.
Due to the spatial resolution of the pressure transducers, it was challenging to assess the hydraulic gradient in the mixture zone in the early stages of the mixing process and in the base layer prior to complete mixing.
Hence, the subsequent analysis in Section~\ref{sec:analysis} focusses on the average hydraulic gradient defined by the head loss through the base and mixture zone.
Generally, the hydraulic gradient was only slightly larger in the base layer compared to the mixture layer, indicating that the average hydraulic gradient was a reasonable parameter to consider in the subsequent analysis.
The porosity distribution indicates that the mixture zone has a lower porosity than the base and filter layer, which is expected given that the base particles are located within the pore space of the filter particles.

At the end of the experiment, a complete mixture zone is formed (Figure~\ref{fig:typical-case}(c)) and fluidisation of the base material within the pores formed by the filter materials can be observed at the top of the cell. 
The average hydraulic gradient is measured across the length of the sample. 
As the Reynolds number is below 1.0 in the base layers prior to fluidisation, the flow regime remained in the laminar range throughout the experiment and Darcy’s law is applicable for the analysis of all tests. 
Due to the size of the filter particles, higher Reynolds number was noted for the filter layer. 
However, the flow rate is controlled by the significantly less permeable base material and mixture zone, and therefore, increased flow resistance caused by turbulence in the filter layer is assumed to be minimal.

\subsection{\label{sub:porosity}Average and Local Porosity Measurements}

No materials were discharged from the coaxial erosion cell during the experiment, and hence, the total solid mass of the particles remained constant throughout the test.
This is beneficial for the determination of the porosity distribution, as any porosity changes in the erosion cell can be solely attributed to the changes in layer heights during the mixing of the base and filter materials.
An average porosity distribution ($n_{ave}$) was obtained using the conventional approach that considered the dry weight of the base and filter particles, along with the layer heights to obtain the volume of each zone \citep{ke2012}. 
This is termed an \emph{average} porosity as it is assumed to be constant across the layer.
$n_{ave}$ is shown in Figure~\ref{fig:typical-case} for the initial condition, partial mixing and complete mixing.
The initial porosity for the base ($n_{ave,B}$) and filter ($n_{ave,F}$) layers in all tests are also listed in Table~\ref{tab:exp-program}.

Using this conventional approach, porosity in the mixture zone was calculated based on the assumption that the base layer below the mixture zone and the filter layer above the mixture zone do not change in porosity during the filtration process.
While visual observations indicated that the base particles below the mixture zone remain at a near-rest condition (confirming the validity of this assumption), the filter particles above the mixture zone experienced some settlement which may be accompanied with slight rearrangement of particles.
While this may result in some changes in porosity, this effect is assumed to be minimal when calculating $n_{ave}$.
Another factor that affects the calculation of the average porosity distribution is the unevenness of the interfaces between layers, and hence, the visual assessment of the layer heights.
It is important to recognise that the layer heights are measured only at the observation window and do not consider any differences within the internal section of the sample. 
In contrast, spatial TDR covered the entire lateral extent of the sample in determining the local porosity profile. 

The local porosity profile ($n_{local}$) obtained along the longitudinal axis of the coaxial erosion cell using spatial TDR enabled detailed insights into the spatial and temporal evolution of porosity during the mixing process.
Figure~\ref{fig:typical-case} compares the local porosity profile with the average porosity profile obtained using the conventional approach outlined above.
A close agreement can be observed between both methods.
A key limitation of the conventional approach is the inability to assess the transient evolution of porosity within the base and filter layer during the mixing process.
This can be quantified using spatial TDR data.
Table~\ref{tab:base-filter-porosity} lists the measured minimum and maximum porosity after the onset of the mixing process by averaging the spatial TDR data across the base and filter layer. 
The minimal variation observed in all cases demonstrates that the assumptions in the conventional approach to calculating $n_{ave}$ are reasonable. 
Only the BI\_FB\_H2 case shows some variation in the porosity of the filter during the mixing process, which is attributed to the significant settlement observed in this test (as detailed in the subsequent analysis shown in Figure~\ref{fig:BI_FB_H2}).
This can also be seen in Figure~\ref{fig:typical-case}(b), where a slight increase in the porosity of the filter layer is noted due to the settlement of the filter particles.
These observations reinforce the ability of spatial TDR to monitor the transient evolution of porosity during the mixing process in filtration experiments, in a manner not possible using conventional approaches to determining porosity evolution.

\begin{table}[ht]
    \caption{\label{tab:base-filter-porosity}Local porosity variation in base and filter layer after onset of mixing process}
    \centering
    \begin{tabular}{|c|c|c|c|c|}
        \hline
         & \multicolumn{2}{c|}{Base Layer} & \multicolumn{2}{c|}{Filter Layer}\\
        \hline
        Test & Min. $\overline{n}_{local}$ & Max. $\overline{n}_{local}$ & Min. $\overline{n}_{local}$ & Max. $\overline{n}_{local}$ \\
        \hline
        \hline
        BI\_FA\_H1 & 0.37 & 0.38 & 0.39 & 0.40 \\
        BI\_FB\_H1 & 0.37 & 0.38 & 0.36 & 0.37 \\
        BII\_FA\_H1 & 0.35 & 0.36 & 0.38 & 0.39 \\
        BII\_FB\_H1 & 0.37 & 0.37 & 0.39 & 0.40 \\
        \hline
        BI\_FA\_H2 & 0.35 & 0.38 & 0.39 & 0.41 \\
        BI\_FB\_H2 & 0.38 & 0.39 & 0.38 & 0.43 \\
        BII\_FA\_H2 & 0.35 & 0.37 & 0.39 & 0.42 \\
        BII\_FB\_H2 & 0.36 & 0.37 & 0.39 & 0.41 \\
        \hline
    \end{tabular}
\end{table}

An important feature of the local porosity profile obtained from spatial TDR is the smooth transitions at the interfaces between layers.
This can be observed in Figure~\ref{fig:typical-case}(b) at the interface of the base material and the mixture zone, as well as the interface between the filter material and the mixture zone.
It can also be observed in Figure~\ref{fig:typical-case}(c) at the interface between the complete mixture zone and the column of water above the sample.
This smooth transition is attributed to the rise time or steepness of the input TDR signal. 
As the steepness of the signal increases, smaller transitions may be detected by spatial TDR.
However, steeper input signals also perturb the measured TDR traces in such a way that the result of the inversion is adversely effected by the presence of large oscillations in the target parameters. 
Following a parametric investigation, a rise time of 1000 ns was found to provide a suitable trade-off between spatial sensitivity in the identification of transitions and the clarity of porosity distributions obtained from the inversion technique.
The subsequent analysis will demonstrate that despite the smoothness of the porosity profile, the transitions between layers can be readily distinguished with a reasonable degree of accuracy.

\begin{figure}
    \centering
    \includegraphics[page=1, trim=0cm 0cm 0cm 0cm, clip, width=0.85\textwidth]{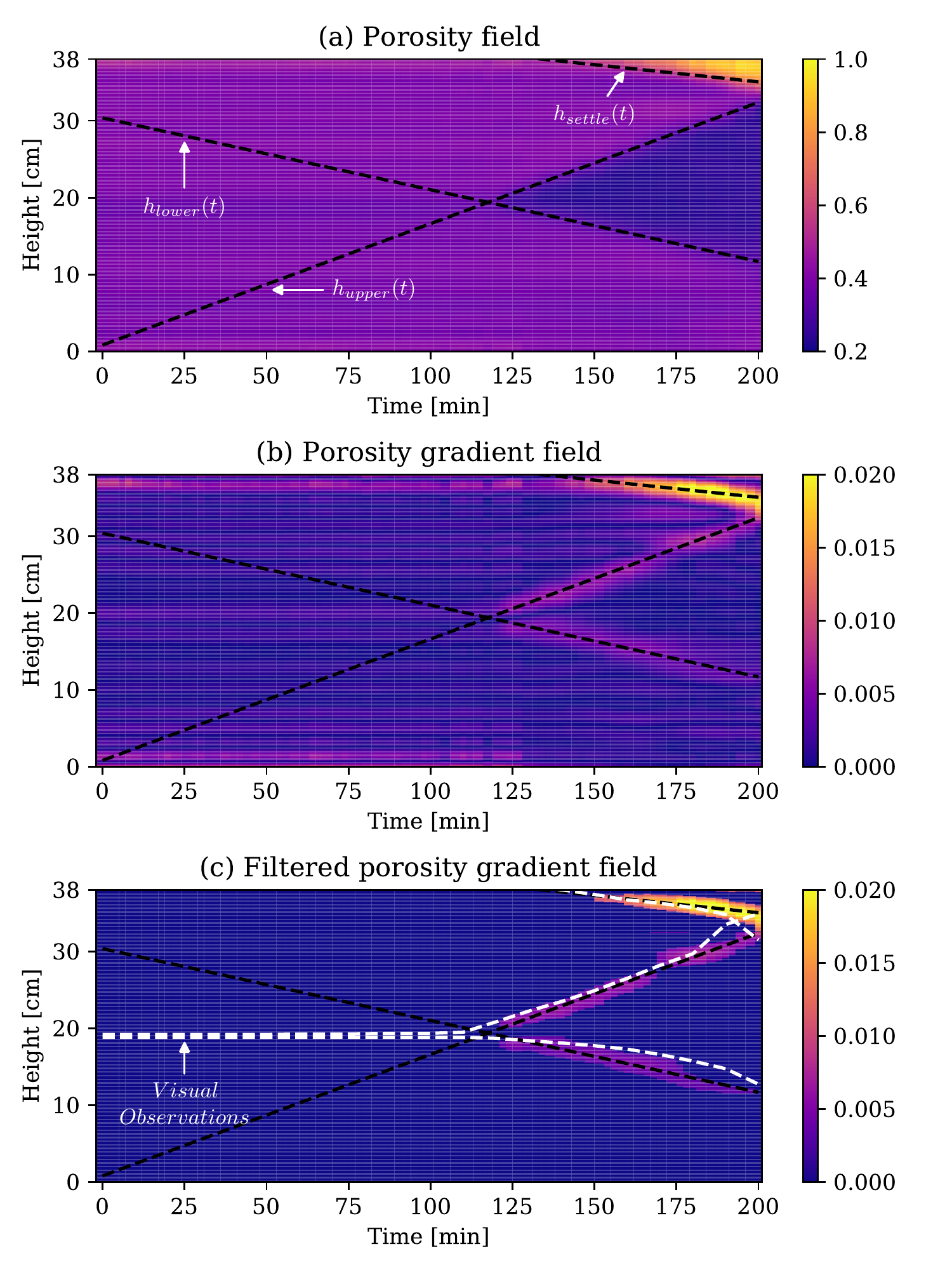}
    \caption{\label{fig:porosity-map}(a) Porosity field map, with best-fit lines for $h_{lower}\left(t\right)$, $h_{upper}\left(t\right)$ and $h_{settle}\left(t\right)$. (b) Porosity gradient field map with the interfaces between zones more clearly distinguished. (c) Map of the filtered porosity gradient field, with the visual observations of the mixture zone (i.e. the lower and upper limits of the mixture zone) shown as white dashed line.}
\end{figure}

The transient evolution of the local porosity profile is graphically displayed in the porosity field map shown in Figure~\ref{fig:porosity-map}, where the colour indicates the local porosity obtained from spatial TDR at a given height along the longitudinal axis of the coaxial erosion cell and at a given time.
The formation of a mixture zone is clearly visible in the porosity field map shown in Figure~\ref{fig:porosity-map}(a), as indicated by the darker shaded area (signifying a decrease in $n_{local}$).
In addition, the settlement of the filter layer is also visible in Figure~\ref{fig:porosity-map}(a) by the lighter shaded area (signifying an increase in $n_{local}$) in the top-right hand corner of the map. 
From the porosity field map, three important characteristics were identified:
\begin{enumerate}
    \item Lower limit of the mixture zone, $h_{lower}\left(t\right)$;
    \item Upper limit of the mixture zone, $h_{upper}\left(t\right)$; and
    \item Settlement line of the filter layer, $h_{settle}\left(t\right)$
\end{enumerate}
Note that all three characteristic heights are defined from the reference datum at the base of the observation window.
The intersection of $h_{lower}\left(t\right)$ and $h_{upper}\left(t\right)$ provides the basis for the onset of filtration, while the gradients of $h_{lower}\left(t\right)$ and $h_{upper}\left(t\right)$ are indicative of the rate of progression of erosion during filtration.
The rate of settlement of the filter can be inferred from the gradient of $h_{settle}\left(t\right)$, as well as the final height of the sample after complete mixing.
Therefore, a quantitative description of the mixing process from onset through to progression can be obtained from the porosity field map shown in Figure~\ref{fig:porosity-map}.
The porosity field map demonstrates that the boundaries between layers can be distinguished despite the smoothness of the local porosity profile, suggesting that the inversion process leads to an approximate spatial resolution of a few centimetres for the local porosity profile.

In order to quantify $h_{lower}\left(t\right)$, $h_{upper}\left(t\right)$ and $h_{settle}\left(t\right)$, the absolute gradient of the porosity field was considered.
This is shown in Figure~\ref{fig:porosity-map}(b).
Note that the gradient map only considers the longitudinal gradient along the sample height (the gradient in time is not considered).
As shown in Figure~\ref{fig:porosity-map}(b), the gradient map more clearly identifies the interfaces between the base and the mixture zone (i.e. $h_{lower}\left(t\right)$), as well as the interface between the filter and the mixture zone (i.e. $h_{upper}\left(t\right)$). 
In addition, the settlement line of the filter layer (i.e. $h_{settle}\left(t\right)$) is also apparent in Figure~\ref{fig:porosity-map}(b).
To better distinguish these characteristic lines, the gradient map is filtered by setting those points with an absolute gradient less than some threshold value to zero.
The filtered porosity gradient map is shown in Figure~\ref{fig:porosity-map}(c).
Different threshold values where considered for each of the characteristic lines by dividing the porosity gradient map into three manually adjusted horizontal zones.
From the filtered porosity gradient map, a line of best-fit was considered to define $h_{lower}\left(t\right)$, $h_{upper}\left(t\right)$ and $h_{settle}\left(t\right)$.
The use of the three horizontal zones and corresponding different threshold values for filtering ensured that each of the best-fit lines where not influenced by neighbouring zones (i.e. upper segment of $h_{upper}\left(t\right)$ was not affected by the lower segment of $h_{settle}\left(t\right)$.
These lines of best-fit are shown in Figure~\ref{fig:porosity-map} by the black dashed lines. 
In Figure~\ref{fig:porosity-map}(c), the visual observations of the mixture zone are also shown as white dashed lines.
A good agreement can be observed for the visual observations and the lines of best-fit for $h_{lower}\left(t\right)$, $h_{upper}\left(t\right)$ and $h_{settle}\left(t\right)$.

\subsection{\label{sub:onset}Limiting Onset Condition}

\begin{figure}
    \centering
    \includegraphics[page=1, trim=0cm 0cm 0cm 0cm, clip, width=0.85\textwidth]{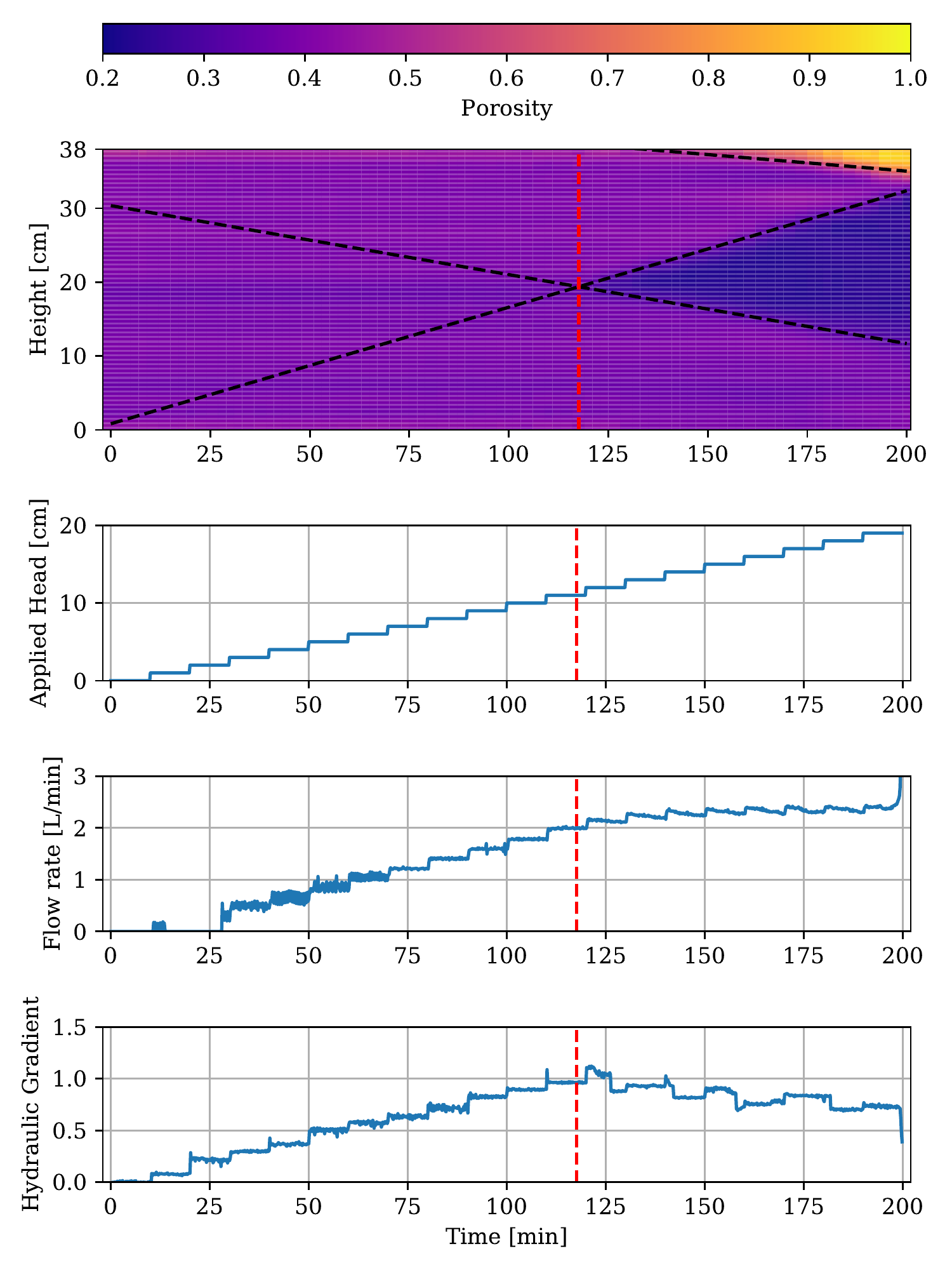}
    \caption{\label{fig:BI_FA_H1}Porosity field map and experimental measurements for BI\_FA\_H1. The lower and upper limit of the mixing zone is delineated, as well as the settlement line, as per Section~\ref{sub:porosity}. The limiting onset condition is shown by the red dashed line.}
\end{figure}

\begin{figure}
    \centering
    \includegraphics[page=1, trim=0cm 0cm 0cm 0cm, clip, width=0.85\textwidth]{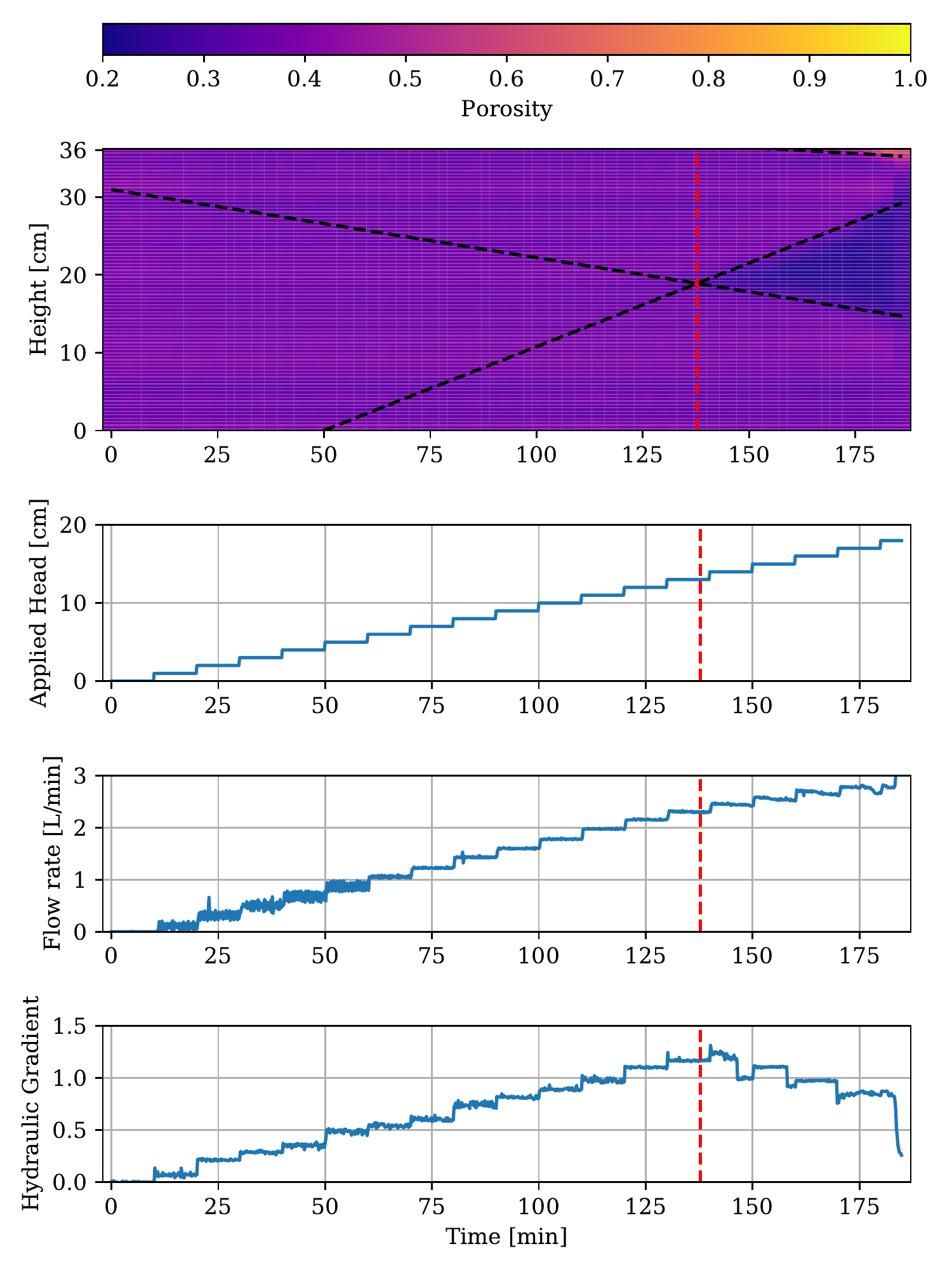}
    \caption{\label{fig:BI_FB_H1}Porosity field map and experimental measurements for BI\_FB\_H1.}
\end{figure}

\begin{figure}
    \centering
    \includegraphics[page=1, trim=0cm 0cm 0cm 0cm, clip, width=0.85\textwidth]{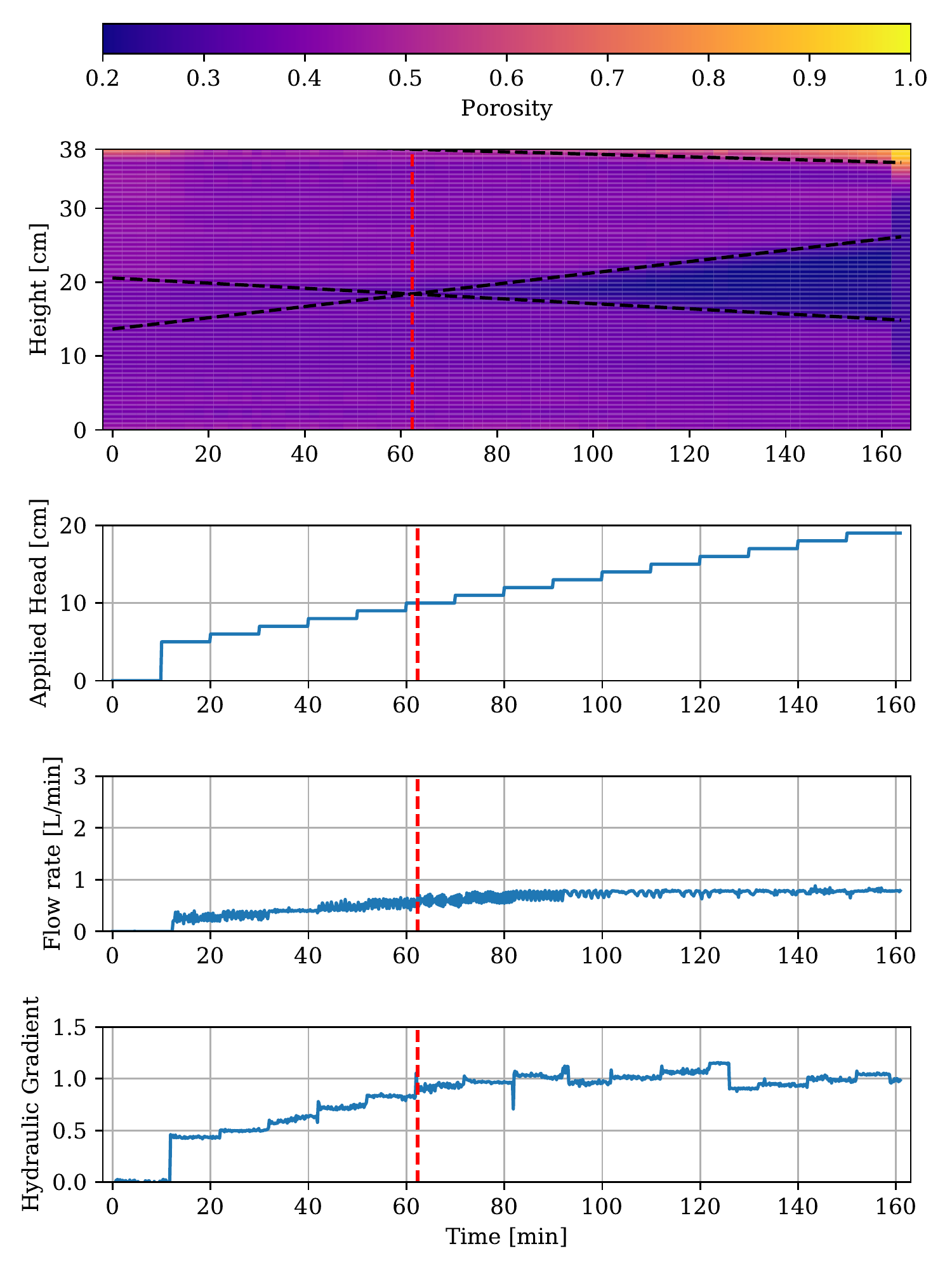}
    \caption{\label{fig:BII_FA_H1}Porosity field map and experimental measurements for BII\_FA\_H1.}
\end{figure}

\begin{figure}
    \centering
    \includegraphics[page=1, trim=0cm 0cm 0cm 0cm, clip, width=0.85\textwidth]{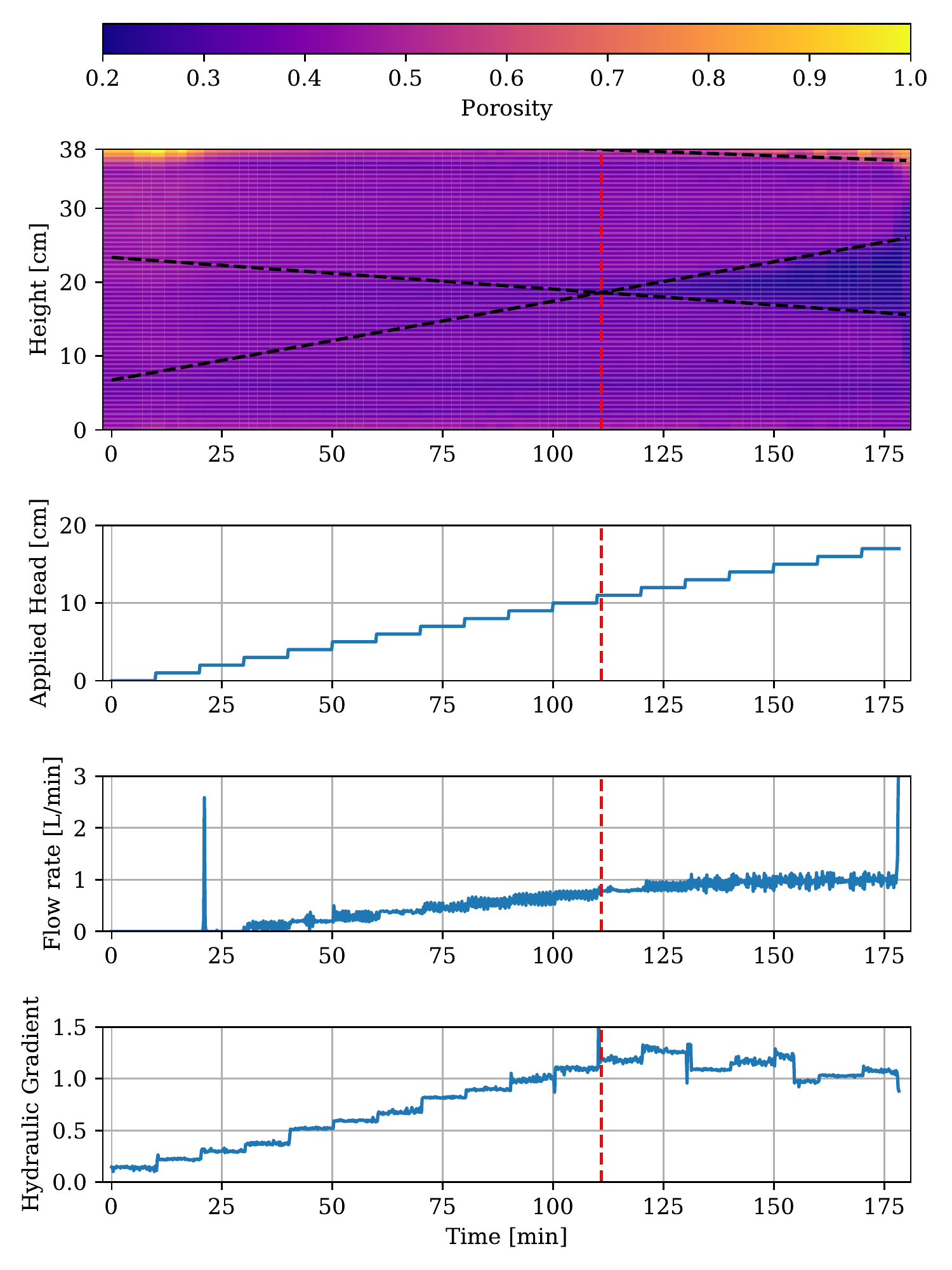}
    \caption{\label{fig:BII_FB_H1}Porosity field map and experimental measurements for BII\_FB\_H1.}
\end{figure}

The onset of the mixing process in filtration experiments occurs when the driving fluid force on the base particles exceeds the counter-acting gravitational force and any inter-particle contact forces. 
At the limiting onset condition, the mixing process is about to commence.
However, the base particles may not necessarily be transported into the filter layer at this stage. 
While it is challenging to observe this limiting onset condition, it can be inferred from the porosity field map.
By simultaneously solving $h_{lower}\left(t\right)$ and $h_{upper}\left(t\right)$, the time at which the mixture zones starts to form ($t_{onset}$) can be determined along with the height along the sample at which the mixture forms ($h_{onset}$). 
The porosity field maps for all cases in Table~\ref{tab:exp-program} with hydraulic boundary condition H1 are shown in Figures~\ref{fig:BI_FA_H1}-\ref{fig:BII_FB_H1}.
The onset condition was more readily observed for H1. 
Figures~\ref{fig:BI_FA_H1}-\ref{fig:BII_FB_H1} also includes the profile of the applied hydraulic head, the measured flow rate and, the average hydraulic gradient ($i_{ave}$) across the base and mixture layers.
Prior to the onset condition, the average hydraulic gradient is given by the hydraulic gradient in the base layer.

\begin{table}[b]
    \caption{\label{tab:onset}Key Characteristics at Limiting Onset Condition}
    \centering
    \begin{tabular}{|c|c|c|c|c|c|c|}
        \hline
        Test & $t_{onset}$ & $h_{onset}$ & $q_{onset}$ & $i_{onset}$ & $i_{crit}^{Terzaghi}$ & $i_{crit}^{Ziems}$\\
        & [min] & [cm] & [L/min] &  &  & \\
        \hline
        \hline
        BI\_FA\_H1  & 117.59 & 19.38 & 2.00 & 0.96 & 0.93 & 0.78 \\ 
        BI\_FB\_H1  & 137.82 & 18.90 & 2.30 & 1.17 & 0.95 & 0.88 \\ 
        BII\_FA\_H1 & 62.29  & 18.40 & 0.46 & 0.93 & 0.96 & 0.78 \\ 
        BII\_FB\_H1 & 110.94 & 18.59 & 0.77 & 1.18 & 0.96 & 0.88 \\ 
        \hline
    \end{tabular}
\end{table}

$t_{onset}$ is shown as a red dashed line in Figures~\ref{fig:BI_FA_H1}-\ref{fig:BII_FB_H1} and is listed in Table~\ref{tab:onset}.
$h_{onset}$ is also listed in Table~\ref{tab:onset} and shows a close agreement with the visual observations noted in Table~\ref{tab:exp-program} for the initial layer heights.
For the visual observations, onset was identified slightly earlier compared to $t_{onset}$ obtained from the porosity field map.
This is expected because the observation window is located on the walls of the coaxial cell where particles can migrate with greater ease.
The small amount of particle movements that are observable through the observation window are too small to be detected by spatial TDR. 
Nevertheless, the porosity field map produced from the spatial TDR measurements provides an accurate estimate of the limiting onset condition across the entire lateral extent of the sample.

The flow rate at the limiting onset condition, $q_{onset}$, is listed in Table~\ref{tab:onset}, from which the critical seepage velocity can be determined by considering the area of the annulus formed by the inner and outer conductor of the coaxial cell and the average porosity of the base layer.
$q_{onset}$ exhibited a strong dependence on the size of the base particles, with a higher $q_{onset}$ observed for larger base particles (Base I comprised larger particles than Base II).
In addition, a slight dependence on the size of the filter particles is observed with a higher $q_{onset}$ noted for smaller filter particles (Filter B comprised smaller particles than Filter A).

The hydraulic gradient in the base layer at the limiting onset condition, $i_{onset}$, is listed in Table~\ref{tab:onset}.
$i_{onset}$ shows a strong dependence on the size of the filter particles, with a lower $i_{onset}$ noted for the larger filter particles in Filter A.
The larger filter particles in Filter A provides a larger pore opening for the base particles to be transported through, and hence, the movement of the base particles are less influenced by the presence of the overlying filter particles leading to a lower $i_{onset}$. 
For this condition, $i_{onset}$ would be expected to approach the Terzaghi critical hydraulic gradient, which is listed in Table~\ref{tab:onset}, and calculated based on the average porosity of the base layer and a specific gravity of $G_s = 2.5$ for the glass beads.
In addition, the larger mass of the filter particles in Filter A can lead to partial bearing failure at the contact point between the filter particles and the base layer. 
This is a result of the reduced effective stress at the base-filter interface due to upward seepage flow, leading to local settlement of the filter particles.
As the filter particle experiences local settlement, they are embedded within the base layer, thereby supporting the mass of the filter particle over a larger extent. 
In this scenario, filter particles can experience further partial bearing failure when the adjacent base particles are transported by seeping water, suggesting that the mixing process is a combination of the transport of base particles and the settlement of filter particles, which is explored further in Section~\ref{sub:progression}. 
Minimal dependence on the size of the base particles is noted for $i_{onset}$.
The trends observed for $i_{onset}$ are in line with the trends noted in \citet{ziems1969}, although the observed $i_{onset}$ in this study are larger than the critical hydraulic gradients stated in \citet{ziems1969} (listed in Table~\ref{tab:onset}).
Moreover, by considering the trends in $i_{onset}$ and $q_{onset}$ for the four different base-filter combinations, it is clear that the limiting onset condition in these filtration experiments are influenced by hydraulic and geometric effects. 
For the case of contact erosion, \citet{brauns1985} considered the quantity $\kappa = n_F \cdot d_{15F}/d_{85B}$ and stated that for $3 < \kappa < 10$, geometric and hydraulic effects influenced the critical velocity for contact erosion. 
For all tests conducted in this study, $3 < \kappa < 10$, and hence, a similar argument is supported for the case of filtration. 

\subsection{\label{sub:progression}Progression of the Mixture Zone in Filtration Experiments}

\begin{figure}
    \centering
    \includegraphics[page=1, trim=0cm 0cm 0cm 0cm, clip, width=0.85\textwidth]{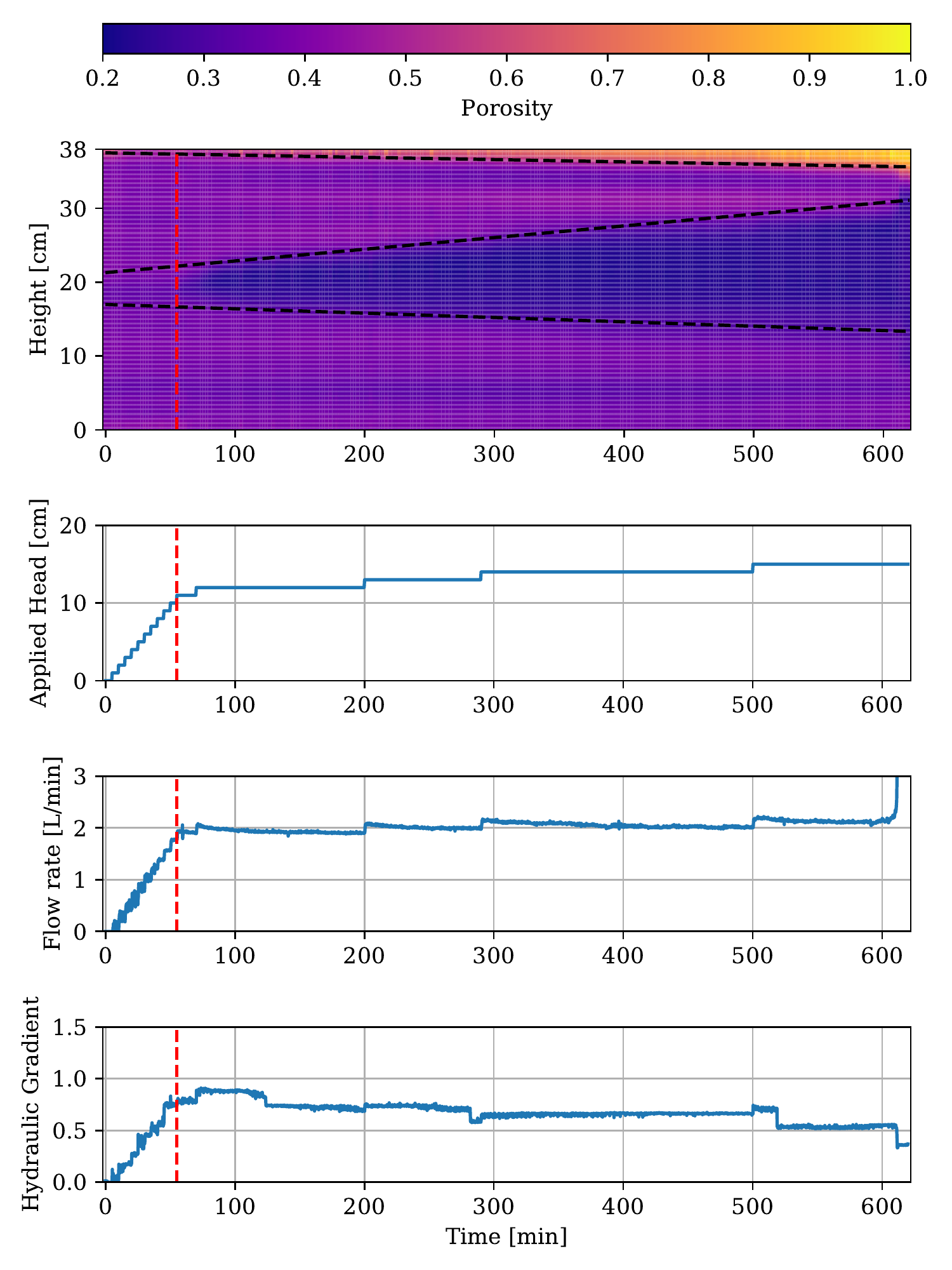}
    \caption{\label{fig:BI_FA_H2}Porosity field map and experimental measurements for BI\_FA\_H2.}
\end{figure}

\begin{figure}
    \centering
    \includegraphics[page=1, trim=0cm 0cm 0cm 0cm, clip, width=0.85\textwidth]{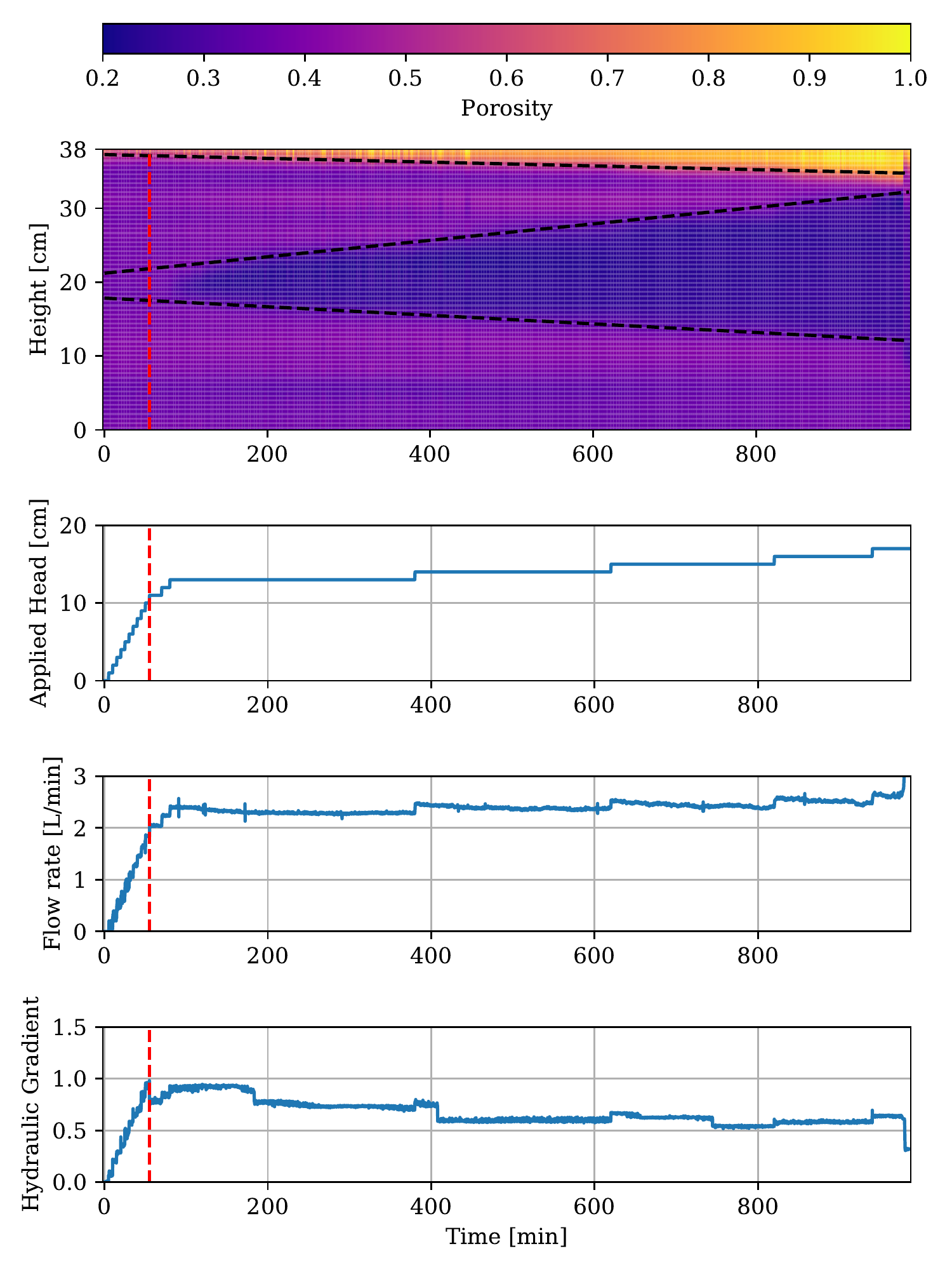}
    \caption{\label{fig:BI_FB_H2}Porosity field map and experimental measurements for BI\_FB\_H2.}
\end{figure}

\begin{figure}
    \centering
    \includegraphics[page=1, trim=0cm 0cm 0cm 0cm, clip, width=0.85\textwidth]{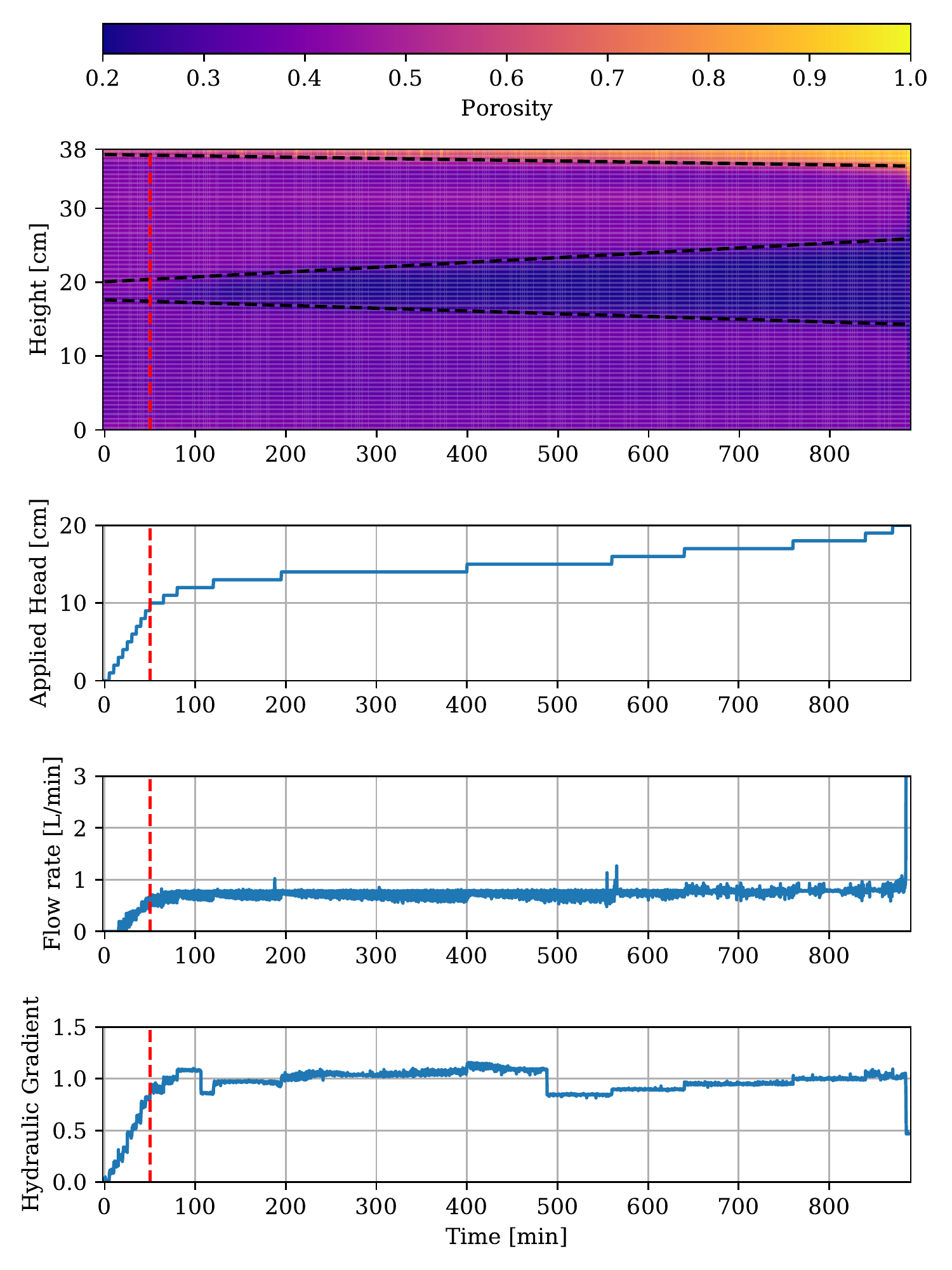}
    \caption{\label{fig:BII_FA_H2}Porosity field map and experimental measurements for BII\_FA\_H2.}
\end{figure}

\begin{figure}
    \centering
    \includegraphics[page=1, trim=0cm 0cm 0cm 0cm, clip, width=0.85\textwidth]{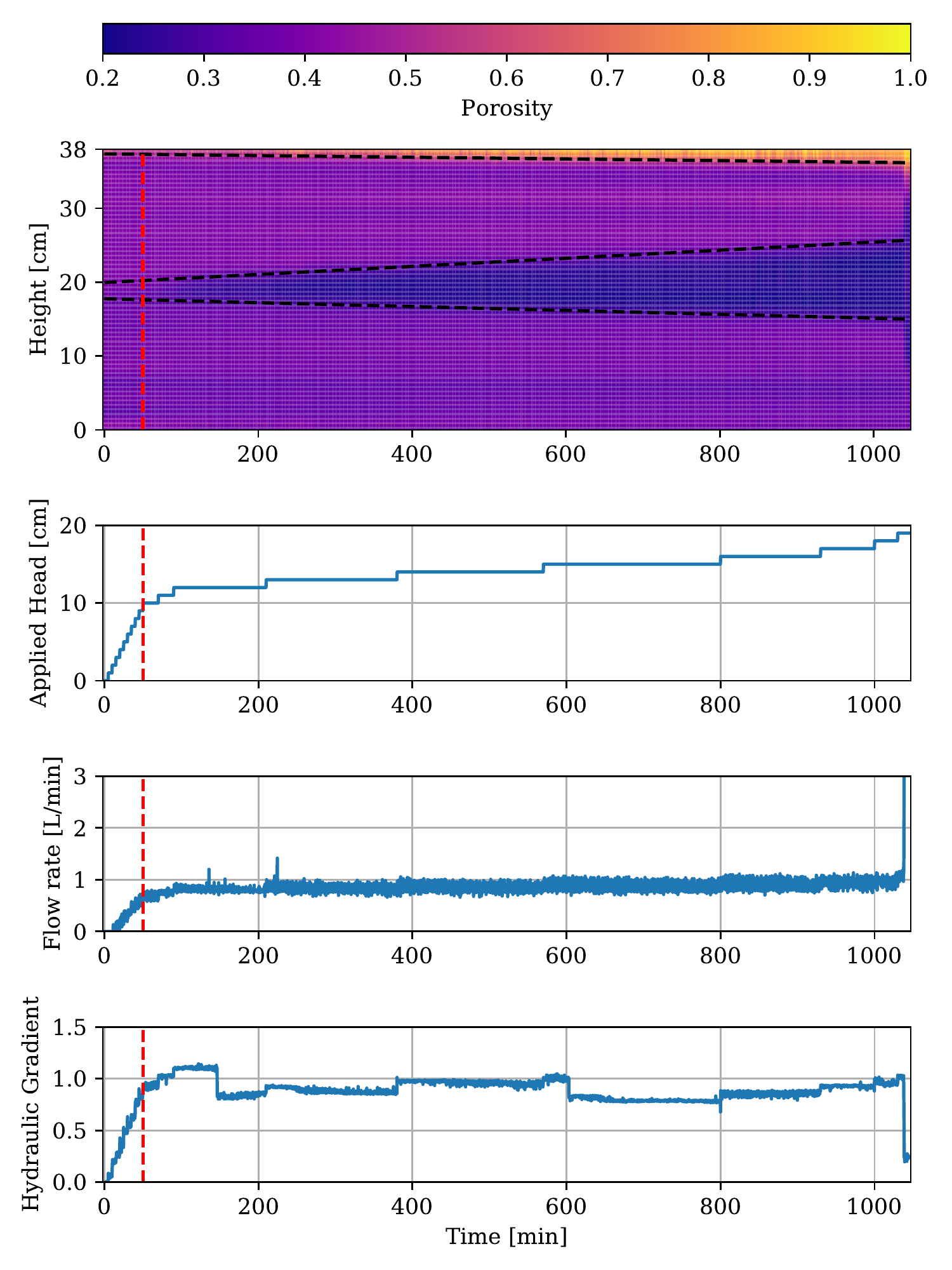}
    \caption{\label{fig:BII_FB_H2}Porosity field map and experimental measurements for BII\_FB\_H2.}
\end{figure}

The progression of erosion in filtration experiments is characterised by the formation of a mixture zone that eventually leads to the complete mixture of the base and filter layers.
By considering the porosity field map, the characteristics of the mixture zone can be quantitatively characterised by considering the gradients of $h_{lower}\left(t\right)$ and $h_{upper}\left(t\right)$.
When considering the onset condition in Section~\ref{sub:onset}, only the tests conducted with the H1 hydraulic boundary condition were considered. 
Characterisation of progression will consider both hydraulic boundary conditions. 
The porosity field maps for tests with H1 have already been presented in Figures~\ref{fig:BI_FA_H1}-\ref{fig:BII_FB_H1}.
The porosity field maps for all cases in Table~\ref{tab:exp-program} with H2 hydraulic boundary condition are shown in Figures~\ref{fig:BI_FA_H2}-\ref{fig:BII_FB_H2}, along with the profile of the applied head, the measured flow rate and average hydraulic gradient ($i_{ave}$) in the base and mixture zones.
Following the onset of the mixing process, the flow rate was observed to remaining approximately constant (or slightly increase) for all cases, with the only exception being the BI\_FB\_H1, while $i_{ave}$ generally decreased due to the increasing length of the base and mixture zone. 
The reduction in $i_{ave}$ was more pronounced in the Base I cases, reflecting the development of a larger mixture zone, which is discussed further below with reference to the porosity field maps. 
Note that abrupt changes in $i_{ave}$ that do not coincide with changes in the applied head are a result of additional pressure transducers entering the mixture zone, thereby adjusting the estimate of $i_{ave}$.

\begin{table}[ht]
    \caption{\label{tab:progression}Characteristics of the Progression of the Mixture Zone}
    \centering
    \begin{tabular}{|c|c|c|c|c|c|}
        \hline
        Test & $m_{upper}$ & $m_{lower}$ & $m_{mix}$ & $m_{settle}$ & $h_{final}$ \\
         & [cm/min] & [cm/min] & [cm/min] & [cm/min] & [cm] \\ 
        \hline
        \hline
        BI\_FA\_H1 & 0.1577 & -0.0934 & 0.2511 & -0.0448 & 35.0 \\
        BI\_FB\_H1 & 0.2148 & -0.0875 & 0.3024 & -0.0312 & 35.2 \\
        BII\_FA\_H1 & 0.0762 & -0.0347 & 0.1109 & -0.0176 & 36.2 \\
        BII\_FB\_H1 & 0.1067 & -0.0429 & 0.1496 & -0.0215 & 36.5 \\
        \hline
        BI\_FA\_H2 & 0.0158 & -0.0059 & 0.0217 & -0.0031 & 35.6 \\
        BI\_FB\_H2 & 0.0111 & -0.0058 & 0.0169 & -0.0025 & 34.7 \\
        BII\_FA\_H2 & 0.0065 & -0.0037 & 0.0103 & -0.0017 & 35.7 \\
        BII\_FB\_H2 & 0.0055 & -0.0026 & 0.0081 & -0.0011 & 36.2 \\
        \hline
    \end{tabular}
\end{table}

From the porosity field maps for H2, it is clear that the best-fit lines provide an accurate representation of the progression of the mixture zone, but comes at the cost of being unable to identify the onset condition. 
This is due to the use of linear best-fit curves and the longer time-scales considered in these tests. 
Therefore, the onset condition in H2 tests was established using visual observations and marked in Figures~\ref{fig:BI_FA_H2}-\ref{fig:BII_FB_H2} by the red dashed line.
Nevertheless, characteristics of the progression of the mixture zone was accurately captured in all tests.
The key characteristics are the gradient $m_{lower}$ of the lower limit of the mixture zone, $h_{lower}\left(t\right)$, and the gradient $m_{upper}$ of the upper limit of the mixture zone, $h_{upper}\left(t\right)$.
For all tests considered, $m_{lower}$ and $m_{upper}$ are listed in Table~\ref{tab:progression}, noting that the gradients are in units of cm/min.
In addition, $m_{mix} = m_{upper}-m_{lower}$ is also listed in Table~\ref{tab:progression} and is a measure of the rate at which the mixture zone developed during filtration.
$m_{mix}$ is approximately an order of magnitude larger in H1 tests compared to tests conducted under H2, indicating that the mixing rate is dependent on the rate of hydraulic loading.
The same observations are noted for $m_{lower}$ and $m_{upper}$.
This is expected as the applied head was increased at a higher rate for H1 (1 cm increments every 10 minutes) compared to H2, where the applied head was kept constant until no erosion was visibly observed and then increased by 1 cm.
This can be visualised by comparing the applied head profiles in Figures~\ref{fig:BI_FA_H1}-\ref{fig:BII_FB_H1} for H1 and Figures~\ref{fig:BI_FA_H2}-\ref{fig:BII_FB_H2} for H2.

Another key observation of the mixture zone is that $\left| m_{upper} \right| > \left| m_{lower} \right|$ for all tests with $1.5 < \left| \frac{m_{upper}}{m_{lower}} \right| < 3.0$.
This is indicative of the nature of the mixing process. 
While it is typically assumed that base particles are transported under seepage flows, this is not the only mechanism at play when considering the formation of the mixture zone. 
Accompanying the transport of base particles is the settlement of the filter particles into the base layer. 
This is a result of partial bearing failure locally where the filter particles touch the base layer, as the effective stress reduces at the base-filter interface within increasing applied head for upward flow. 
The rate of settlement of the filter particles into the base layer is captured through $m_{lower}$.
Irrespective of the hydraulic boundary condition, $m_{lower}$ is primarily dependent on the size of the base particles, with $m_{lower}$ increasing with increasing base particle size (as Base I comprised larger particles).
When the filter particles settle in to the base layer, the neighbouring base particles are displaced upwards, and this process in addition to the transport of base particles by upward seepage flow is captured by $m_{upper}$.
While these processes cannot be readily distinguished in the $m_{upper}$ parameter, it provides justification for $\left| m_{upper} \right| > \left| m_{lower} \right|$.
Similarly to observations for $m_{lower}$, $m_{upper}$ was primarily dependent on the size of the base particles, with $m_{upper}$ larger for tests with Base I.
The implications of these trends in $m_{lower}$ and $m_{upper}$, is that the rate of growth in the mixture zone was higher for all tests with Base I.
This is evident when comparing the porosity field maps, with a larger mixture zone noted in all tests with Base I, irrespective of the size of the filter or the hydraulic boundary condition.
It is hypothesised that this observation is due to increased frequency of collisions between the smaller base particles for the Base II cases, thereby leading to greater energy dissipation for the base particles and a smaller mixture zone.
However, inter-particle collisions of the base particles cannot be readily observed or measured in these experiments, and hence, further research is required to confirm this hypothesis. 
A potential approach is to consider numerical simulations of fluid-particle systems, such as coupled computational fluid dynamics and discrete element method, to probe the inter-particle collisions during the mixing process.

In addition to the formation of the mixture zone, the progression of filtration also leads to the settlement of the sample, which is characterised by the settlement line, $h_{settle}\left(t\right)$.
The gradient of the settlement line, $m_{settle}$, reflects the rate of settlement and is listed in Table~\ref{tab:progression}.
Similarly to the observations noted above for $m_{mix}$, $m_{settle}$ is an order of magnitude larger for the H1 tests due to the higher rate of hydraulic loading. 
In addition, $m_{settle}$ is larger in the tests with Base I, indicating slightly larger total settlement of the sample and this is reflected in the final height of the sample, $h_{final}$, as listed in Table~\ref{tab:progression}.
Nevertheless, it is noted that similar $h_{final}$ was observed for each base-filter combination irrespective of the hydraulic boundary condition, suggesting that the final height of the complete mixture is independent of the hydraulic loading path.
It is important to note that the complete mixture formed very quickly, and hence, the lower and upper limits of the mixture zone may not necessarily coincide with the complete mixture condition.

\section{\label{sec:conclusions}Conclusions}

This study has demonstrated the ability of spatial Time Domain Reflectometry (spatial TDR) to accurately make physical observations of the transient evolution of the porosity distribution during the mixing process in filtration experiments. 
This was achieved by using a purpose-built permeameter that acted as a coaxial transmission line enabling electromagnetic measurements in the form of spatial TDR, from which the longitudinal porosity distribution was obtained. 
The filtration experiments conducted in this study investigated the influence of the size of base and filter particles, in addition to the influence of the rate of hydraulic loading by considering two different hydraulic boundary conditions. 
The data obtained from spatial TDR enabled the construction of a porosity field map that showed the spatial variation of porosity along the longitudinal axis of the permeameter, as well as the temporal variation of porosity during the mixing process of the base and filter particles. 
From the porosity field map, three key characteristics were quantitatively described using lines of best-fit, including the lower limit of the mixture zone, the upper limit of the mixture zone and the settlement line of the sample.  
Using these key characteristics of the transient evolution of the porosity distribution, the onset and progression of the mixing process in filtration experiments was quantitatively investigated. 

The limiting onset condition is clearly visible by the intersection of the lower and upper limits of the mixture zone on the porosity field map, providing the time at which mixing commenced, from which the critical flow rate (and hence, critical seepage velocity) and critical hydraulic gradient at the limiting onset condition could be determined.
The critical flow rate showed a strong dependence on the size of the base particles, with the critical flow rate increasing for larger base particles. 
In contrast, the critical hydraulic gradient exhibited a stronger dependence on the size of the filter particles, with a lower critical hydraulic gradient observed with increasing filter particle size.
Based on the observations at the limiting onset condition, it was evident that for the tests considered in this study, both geometric and hydraulic factors affected the onset condition. 

The transient evolution of porosity during the mixing process was also quantitatively captured by the key characteristics of the porosity field map, thereby enabling detailed insights into the progression of erosion. 
A key observation was that the formation of the mixture zone was influenced by two mechanisms: (i) the transport of base particles into the filter layer due to upward seepage flows; and (ii) the settlement of the filter particles into the base layer due to the reduction of the effective stress at the base-filter interface leading to partial bearing failure. 
Both mechanisms could be inferred from spatial TDR data by considering the differing gradients of the lower and upper limits of the mixture zones from the porosity field map. 
These gradients also provided quantitative insights into the progression characteristics. 
By considering two different hydraulic boundary conditions, it was demonstrated that the rate of hydraulic loading influences the rate at which the mixture zone developed. 
In addition, the rate of development of the mixture zone was significantly higher for all tests conducted with larger base particle sizes, indicating the strong dependence of the progression process on the size of the base particles. 
Concurrent with these observations on the mixture zone, the settlement of the sample was also shown to be dependent on the base particle size, with larger settlements noted in the tests with the larger base particles. 
Furthermore, the final height of the sample after complete mixture of the base and filter particles was insensitive to the hydraulic boundary condition. 

These observations on the transient evolution of the porosity distribution demonstrates the capability of spatial TDR to make physical observations of the mixing process in filtration experiments from onset to progression.
Ongoing research aims to employ these physical observations for calibration of numerical particle-scale simulations of the filtration process using coupled computational fluid mechanics and discrete element method \citep{smith2020,che2021} and pore network models \citep{van2018,sufian2019}.
These physical observations can also be applied to large deformation models employing smooth particle hydrodynamic or the material point method. 
This study was limited to considering base-filter combinations that displayed continuing erosion, thereby enabling the formation of a complete mixture zone, which highlighted the capability of the approach to investigate the mixing process from onset to progression.
Nevertheless, future research aims to extend the coaxial erosion cell to investigate a wider range of base-filter combinations and different hydraulic boundary conditions, including cyclic hydraulic boundary conditions. 
The coaxial erosion cell presented in this study is also not limited to investigating filtration and ongoing research is investigating the onset and progression of gap-graded soils susceptible to suffusion. 
This demonstrates the wide applicability of spatial TDR to investigate a range of processes associated with internal erosion.

\section*{Acknowledgements}
This study was funded by the Australian Research Council (ARC) Discovery Project (DPI120102188), titled 'Hydraulic erosion of granular structures: Experiments and Computational Simulations'.
A. Scheuermann was supported by the Queensland Science Fellowship and an ARC Future Fellowship (FT180100692).
T. Bore was supported by the ARC Discovery Early Career Researcher Award (DE180101441).

\nomenclature{$l_{screen}$}{Initial thickness of the screening layer}
\nomenclature{$l_{base}$}{Initial thickness of the base layer}
\nomenclature{$l_{filter}$}{Initial thickness of the filter layer}
\nomenclature{$l_{mixture}$}{Initial thickness of the mixture layer}
\nomenclature{$n_{ave}$}{Average porosity based on layer heights}
\nomenclature{$n_{ave,B}$}{Initial average porosity of the base layer}
\nomenclature{$n_{ave,F}$}{Initial average porosity of the filter layer}
\nomenclature{$n_{local}$}{Local porosity based on spatial TDR data}
\nomenclature{$h_{lower}\left(t\right)$}{Lower limit of the mixture zone}
\nomenclature{$h_{upper}\left(t\right)$}{Upper limit of the mixture zone}
\nomenclature{$h_{settle}\left(t\right)$}{Settlement line of the filter layer}
\nomenclature{$t_{onset}$}{Time at the limiting onset condition from the porosity field map}
\nomenclature{$h_{onset}$}{Height at which the limiting onset condition occurs from the porosity field map}
\nomenclature{$q_{onset}$}{Flow rate at the limiting onset condition}
\nomenclature{$i_{ave}$}{Average hydraulic gradient across the base and mixture layers}
\nomenclature{$i_{onset}$}{Average hydraulic gradient at the limiting onset condition}
\nomenclature{$m_{lower}$}{Gradient of the lower limit of the mixture zone}
\nomenclature{$m_{upper}$}{Gradient of the upper limit of the mixture zone}
\nomenclature{$m_{mix}$}{Rate of development of the mixture zone}
\nomenclature{$m_{settle}$}{Gradient of the settlement line}
\nomenclature{$h_{final}$}{Final height of the sample after complete mixture}

\printnomenclature

\bibliographystyle{ms}
\bibliography{ms}

\end{document}